\documentclass[aps,prd,showpacs,preprintnumbers,amsmath,amssymb,nofootinbib,11pt]{revtex4}
\usepackage[dvips]{graphicx}
\usepackage[colorlinks,dvipdfm]{hyperref}
\def\N{N_C}
\def\RCT{R\chi T}
\def\t{\tau}
\def\n{\nu}

\def\p{\pi}
\def\CPT{\chi PT}
\newcommand{\be}{\begin{equation}}
\newcommand{\ee}{\end{equation}}
\newcommand{\ba}{\begin{array}}
\newcommand{\bqa}{\begin{eqnarray}}
\newcommand{\eqa}{\end{eqnarray}}
\newcommand{\cO}{{\cal O}}

\newcommand{\strich}[1]{#1  \! \! \! \slash}
\newcommand{\gnr}{\Gamma_{\tau^- \to \nu_\tau P^-}}
\begin{document}
\title{One meson radiative tau decays}
\author{Zhi-Hui~Guo$^1$, Pablo Roig$^2$}

\affiliation{ 1: CAFPE and Departamento de F\'{\i}sica Te\'orica y del
Cosmos, Universidad de Granada,
 Campus de Fuente Nueva, E-18002 Granada, Spain.
\\
2:  Laboratoire de Physique Th\'eorique (UMR 8627)
Universit\'e de Paris-Sud XI, B\^atiment 210,
 91405. Orsay cedex, France.}

\begin{abstract}
We have studied the one-meson radiative tau decays: $\tau^- \to \nu_\tau \pi^-(K^-) \gamma$, computing the structure dependent contributions within
a Lagrangian approach based on the large-$\N$ limit of $QCD$ that ensures the proper low-energy limit dictated by chiral symmetry. Upon imposing the
short-distance $QCD$ constraints to the form-factors we are able to predict the structure dependent radiation without any free parameter and, therefore,
the relevant observables for the decay $\tau^- \to \nu_\tau \pi^- \gamma$: the photon energy spectrum, the invariant mass spectrum of the meson-photon system and the integrated
decay rate. We also discuss the remaining uncertainties in these observables for the $\tau^- \to \nu_\tau K^- \gamma$ decay. According to our results,
the present facilities could detect these rare decays for the first time in the near future allowing for a stringent test of our predictions.
\end{abstract}

\preprint{CAFPE-145/10}
\preprint{UG-FT-275/10}

\pacs{
13.35.Dx,
12.39.Fe,
11.15.Pg
\\
Keywords: Decays of tau lepton;  Chiral Lagrangians;  $1/N_C$ expansion  }
\maketitle
\section{Introduction}\label{Intro}
The decays of the $\t$ lepton are an ideal benchmark to analyze the hadronization of the spin-one $QCD$ currents and to learn about
the treatment of the strong interaction involving the intermediate meson dynamics in rather clean conditions, since the electroweak part of the decay
is under good theoretical control and  the light-flavoured hadron resonances rule these processes
 \cite{EarlyPapers, RecentTalksandReviews, Decker:1992kj, Decker:1993ut, Geng:2003prd, Kuhn:1990ad, Kuhn:1992nz, GomezDumm:2003ku, Kpi, Guo:2008sh, Dumm:2009kj, Dumm:2009va}.

Among the various exclusive tau decays, we are particularly interested in the one meson radiative tau decays in this article, i.e. $\tau \to \pi(K) \nu \gamma$.
Unlike in many of the multi-pseudoscalar tau decay channels, where the excited vector resonances (such as $\rho'\,, {K^*}'$) could play a
significant role due to the kinematics effects, the channels we are going to study here should be less influenced by the excited resonances,
since the lowest vector resonances such as $\rho(770)\,, K^*(892)$, could already decay into $\pi(K) \gamma$ at their on-shell energy regions.
So these radiative decay processes can provide us with an excellent environment to investigate the lowest resonance states. On the experimental side,
these channels have not been detected yet, which makes our analysis more meaningful,
since the current work could serve as a helpful tool for the measurements of these channels in the B and tau-charm Factories in the near future.

 The decay amplitude for the one meson radiative decay of the $\t$ includes an internal bremsstrahlung ($IB$) component, that is given by $QED$,
and thus can be calculated unambiguously to any desired order in perturbation theory. In addition, one has the structure dependent ($SD$) part, dominated by the
effects of the strong interaction. Lorentz symmetry determines that there are two independent structures, the so-called vector and axial-vector form
factors that encode our lack of knowledge of the precise mechanism responsible for hadronization.

One has then to rely on parametrizations of these form factors. Some of the earliest attempts can be found in Refs.~\cite{EarlyPapers}.
The so-called K\"uhn-Santamar\'ia model \cite{Kuhn:1990ad} became a very popular approach. For a given meson mode it proceeds as follows: it normalizes
the form factor using the lowest order contribution stemming from the low-energy effective field theory of $QCD$ in the light-flavoured sector: Chiral
Perturbation Theory, $\CPT$ \cite{ChPT, Gasser:1983yg, Gasser:1984gg, Gasser:1984ux}. Then, the form factor is constructed in terms of Breit-Wigner
functions weighted by some unknown parameters
(to be fitted to data) in such a way that it vanishes at infinite transfer of momentum in order to obey a Brodsky-Lepage behaviour
\cite{BrodskyLepage, Lepage:1979zb}. This procedure was successful to describe the $ARGUS$ data on the $\t\to\p\p\p\n_\t$ decays \cite{Albrecht:1986kg}. However,
as the data became more precise \cite{Barate:1998uf} it was shown that there was room for improvement \cite{GomezDumm:2003ku, Dumm:2009va}. Among the
theoretical reasons for this, one finds that in the low-energy limit this model is not consistent with $\CPT$ at $\cO(p^4)$ \cite{GomezDumm:2003ku,
Portoles:2000sr}. Assorted versions of the K\"uhn-Santamar\'ia model have been used for several two- and three-meson tau decays \cite{Decker:1992kj,
KSmodels} and also for the radiative one-meson decays we consider in this article \cite{Decker:1993ut}. In general, these works made following
Ref.~\cite{Kuhn:1990ad} suffer from additional problems, as discussed in Ref.~\cite{Roig:2008ev}. In fact, arbitrary parametrizations
are of little help in order to comprehend the characteristics of the hadronization procedure independently of their eventual success in describing
experimental data. The best procedure in order to investigate this problem should be to build the relevant form factors upon as many $QCD$ features as possible.

The $TAUOLA$ library \cite{Jadach:1993hs} is currently the most relevant tool to analyze tau decay data. Although it only incorporated
the K\"uhn-Santamar\'ia models at first, it was enriched with
parametrizations by the experimental collaborations ALEPH and CLEO -collected in Ref.\cite{Golonka:2003xt}- in order to explain their data.
In the B-factories era, however, it has become evident that this strategy is not adequate and $TAUOLA$ has been opened to the introduction of matrix
elements obtained within other approaches that include more properties stemming from $QCD$. This makes it an excellent tool to exploit the synergies of
Monte Carlo methods and theoretical approaches to better understand the large data samples of high quality obtained at the B-factories ($BaBar$,
$Belle$) and the $\t$-charm factories, such as $BES$-III. This is thus an appropriate benchmark where the results of our
research can be applied.

Although one knows the underlying theory, $QCD$, this kind of study is rather involved since there is no fully analytic way of relating
the final state mesons that are detected to the quark and gluon degrees of freedom of the $QCD$ Lagrangian. Moreover we are in the non-perturbative
regime of the strong interaction ($E\lesssim M_\t\sim1.7$ GeV), so any perturbative treatment within $QCD$ will not be a good approach to the problem.

At very low energies ($E\lesssim M_\rho$, being $M_\rho$ the mass of the $\rho$(770) resonance), the chiral symmetry of massless $QCD$
rules the construction of an effective field theory that allows a perturbative expansion in momenta ($p$) and light quark masses ($m$), as ($p^2,\,m_\pi^2$)/
$\Lambda_\chi^2$, being $\Lambda_\chi\sim 4 \pi F\sim M_\rho$ the chiral symmetry breaking scale. $m_\pi$ is the pion mass and $F$ is its decay constant,
with the normalization of $F = 92.4$ MeV.
This theory is $\CPT$, that drives the hadronization of $QCD$ currents into the lightest multiplet of pseudoscalar mesons, including $\pi$, $K$ and $\eta$
particles. This framework was applied to the two- and three-pion tau decays in Ref.~\cite{Colangelo:1996hs} and it was checked that it provides a right
description of the low-energy data though it was incapable of providing a good parametrization for the rest of the spectrum. This study put forward
whatever structure due to resonance exchange that the form factors may have should match the chiral behaviour in the limit where $\chi$PT applies.
As we brought forward before, the K\"uhn-Santamar\'ia models fail to fulfill that condition already at $\cO(p^4)$ in the chiral expansion \cite{GomezDumm:2003ku,
Portoles:2000sr}.

We shall extend $\CPT$ to energies $E\gtrsim M_\rho$, where the expansion parameter of $\CPT$ is no longer valid. In fact, it is not
known how to develop a dual effective theory of $QCD$ in this region. However, there is a construct that has proven to be useful in this regime shedding
light on the appropriate structure of the Lagrangian theory that we could use. This is yielded by the large-$\N$ limit of $SU(\N)$ $QCD$ \cite{LargeNc}
which introduces the inverse of the number of colours (three in the real world) as the parameter to build the expansion upon. In our context, it amounts
to consider an spectrum of an infinite number of zero-width resonances that interact at tree level through local effective vertices. This frame, as
we will see, can be used to establish a starting point in the study of the resonance region, and consequently in the hadronic decays of the tau lepton
we are dealing with. The setting recalls the r\^ole of the Resonance Chiral Theory ($\RCT$) \cite{Ecker:1988te, Donoghue:1988ed} that can be better
understood in the light of the
large-$\N$ limit of $QCD$ as a theory of hadrons\cite{PerisdRafael, Pich:2002xy}.

The theory, although built upon symmetries guided by the large-$\N$ expansion and reproducing the chiral behaviour in the low-energy region, is still
missing an ingredient of $QCD$. At higher energies ($E\gg M_\rho$), where the light-flavoured continuum is reached, perturbative $QCD$ is the appropriate
tool to deal with the description of interactions, which is given in terms of partons. A well-known feature of form factors of $QCD$ currents is their smooth
behaviour at high energies \cite{BrodskyLepage, Lepage:1979zb}. Then, it is plausible that the form factors match this behaviour above the energy region where the
resonances lie. A complementary approach is given by the study of the operator product expansion ($OPE$) of Green functions of $QCD$ currents that are order parameters
of the chiral symmetry breakdown. Refs. \cite{Moussallam, Knecht:2001xc, RuizFemenia:2003hm, Cirigliano:2004ue, Mateu:2007tr, Ecker:1989yg} have evaluated these Green
Functions within a resonance theory and proceeded to match the outcome to the leading
term of the $OPE$ at high transfers of momenta. As we commented it is crucial to take into account this high-energy information to settle a resonance
Lagrangian that implements as many $QCD$ features as possible, whence the described procedure will help to establish relations among some of the
Lagrangian couplings, and eventually fix others \cite{Ecker:1989yg}.

We will consider the $SD$ description of the processes of $\tau^-\to P^- \gamma \nu_\tau$ ($P=\pi,\,K$)
within the framework of $\RCT$ as introduced in this section and discussed in detail in the following one. These channels
have not been observed yet, which is strange according to the most naive expectations of their decay rates. Clarifying this question is one of the main
motivations of our study.

The relative sign between the $IB$ and $SD$ dependent part motivated an addendum to Ref.~\cite{Decker:1993ut}. This confusion was caused by
the fact that they did not use a Lagrangian approach. In any such approach
this should not be an issue. In order to facilitate any independent check, we define the convention we follow as the one used by $PDG$ \cite{PDG08} in order
to relate the external fields $r_\mu, \ell_\mu$ with the physical photon field
\begin{equation}
r_\mu=\ell_\mu= -e\, Q \,A_\mu\,,
\end{equation}
where $e$ is the electric charge of the positron and the charge matrix of the three light flavour quarks is $Q={\rm diag}(\frac{2}{3}, -\frac{1}{3}, -\frac{1}{3})$.
Determining the relative sign between the structure independent ($SI$) and dependent contributions is an
added interest of our computation.

The $SI$ part of the process has been discussed in \cite{Decker:1993ut}. We will compute the $SD$ part
using the Lagrangians in Sect. \ref{Theo}. The kinematics
and differential decay rates are discussed in Sect. \ref{Kinematics}, as well as the general form of the matrix elements and the spectra.
The structure dependent form factors for each mode are computed in Sects. \ref{SDFFpi} and \ref{SDFFK}. The $QCD$
short-distance constraints are presented in Sect. \ref{QCDconstraints} and the phenomenological implications are detailed in Sect. \ref{Pheno}.
Finally, we summarize and discuss our results in Sect. \ref{Conclusions}.

\section{Theoretical framework}\label{Theo}
The hadronization of the currents that rule semileptonic tau
decays is driven by non-perturbative $QCD$. As mentioned in the Introduction,
our methodology stands on the construction of an action, with the relevant
degrees of freedom, led by the chiral symmetry and the known asymptotic
behaviour of the form factors and Green functions driven by large $N_C$ $QCD$. We
will present here those pieces of the action that are relevant for the
study of  one meson radiative decays of the tau lepton. Hence we
will need to include both even- and odd-intrinsic parity sectors.

The large $N_C$ expansion of $SU(N_C)$ $QCD$ implies that, in the $N_C
\rightarrow \infty$ limit, the study of Green functions of $QCD$ currents and associated form factors can
be carried out through the tree level diagrams of a Lagrangian theory that
includes an infinite spectrum of strictly stable states \cite{LargeNc} \footnote{Since light resonances reach their on-shell peaks
in the energy region spanned by the considered decays, the corresponding off-shell widths, which are energy-dependent, need to be implemented as we do
following Refs. \cite{Dumm:2009va, GomezDumm:2000fz}.}.
Therefore the study of the resonance energy region can be performed by
constructing such a Lagrangian theory. However, it is not
known how to implement an infinite spectrum in a model-independent way.
Moreover, it is well known from the phenomenology that the main role is
always played by the lightest resonances. Accordingly it was suggested in
Refs.~\cite{Ecker:1988te, Donoghue:1988ed} that one can construct a
suitable effective Lagrangian involving the lightest multiplets of resonances
and the pseudo-Goldstone bosons ($\pi$, $K$ and $\eta$). This is
indeed an appropriate tool to handle the hadronic decays of the tau lepton and
the pion form factors \cite{GomezDumm:2003ku,
Kpi, Guo:2008sh, Dumm:2009kj, Dumm:2009va, Guerrero:1997ku, SanzCillero:2002bs, Rosell:2004mn}.
The guiding principle in the construction of such a Lagrangian is
chiral symmetry. When
resonances are integrated out from the theory, i.e. one tries to describe
the energy region below such states ($E \ll M_{\rho}$), the remaining
setting is $\CPT$, reviewed in Refs.~\cite{Ecker:1994gg, Pich:1995bw}.

The very low-energy strong interaction in the light quark sector is known to
be ruled by the $SU(3)_L\otimes SU(3)_R$ chiral symmetry of massless $QCD$
implemented in $\CPT$. The leading even-intrinsic-parity ${\cal O}(p^2)$
Lagrangian, which carries the information of the spontaneous symmetry
breaking of the theory, is~:
\begin{equation} \label{eq:op2}
{\cal L}_{\chi {\rm PT}}^{(2)}=\frac{F^2}{4}\langle u_{\mu}
u^{\mu} + \chi _+ \rangle \ ,
\end{equation}
where
\begin{eqnarray}
u_{\mu} & = & i [ u^{\dagger}(\partial_{\mu}-i r_{\mu})u-
u(\partial_{\mu}-i \ell_{\mu})u^{\dagger} ] \ , \nonumber \\
\chi_{\pm} & = & u^{\dagger}\chi u^{\dagger}\pm u\chi^{\dagger} u\ \
\ \ , \ \ \ \
\chi=2B_0(s+ip) \; \; ,
\end{eqnarray}
and $\langle \ldots \rangle$ is short for the trace in the flavour space.
The Goldstone octet of pseudoscalar fields
\begin{equation}  \label{eq:phi_matrix}
\Phi(x) =
\left(
\begin{array}{ccc}
 \displaystyle\frac{1}{\sqrt 2}\,\pi^0 + \displaystyle\frac{1}{\sqrt
 6}\,\eta_8
& \pi^+ & K^+ \\
\pi^- & - \displaystyle\frac{1}{\sqrt 2}\,\pi^0 +
\displaystyle\frac{1}{\sqrt 6}\,\eta_8
& K^0 \\
 K^- & \bar{K}^0 & - \displaystyle\frac{2}{\sqrt 6}\,\eta_8
\end{array}
\right)
\ ,
\end{equation}
is realized non--linearly into the unitary matrix in the
flavour space
\begin{equation}
u(\varphi)=\exp \left\{ \frac{i}{\sqrt{2}\,F} \Phi(x) \right\} \; \; \; ,
\end{equation}
 which under chiral rotations transforms as
\begin{equation}
u(\varphi)  \to  g_R\, u(\varphi)\, h(g,\varphi)^\dagger
                 = h(g,\varphi)\, u(\varphi)\, g_L^\dagger \; \; ,
\end{equation}
with $g \equiv (g_L,g_R) \, \in \, SU(3)_{\mathrm{L}} \otimes
SU(3)_{\mathrm{R}}$ and $h(g,\varphi)\,\in \, SU(3)_V$. External hermitian
matrix fields $r_{\mu}$, $\ell_{\mu}$, $s$ and $p$ promote the global
$SU(3)_{\mathrm{L}} \otimes SU(3)_{\mathrm{R}}$ symmetry to a local one.
Thus, interactions with electroweak bosons can be accommodated through the
vector $v_{\mu} = (r_{\mu} + \ell_{\mu}) / 2$ and axial--vector $a_{\mu} =
(r_{\mu} - \ell_{\mu}) / 2$ fields. The scalar field $s$ incorporates the
explicit chiral symmetry breaking through the quark masses taking $s = {\cal
M} \, + \ldots$, with ${\cal M} =  \mathrm{diag}(m_u,m_d,m_s)$, where we will always work
in the isospin limit in the present discussion, i.e. $m_u=m_d$. Finally,
at lowest order in the chiral expansion $F$ is
the pion decay constant and $B_0 F^2 = - \langle \bar{\psi}\psi \rangle$, with the quark condensate of
$\langle \bar{\psi}\psi \rangle = \langle \bar{u}u \rangle= \langle \bar{d}d \rangle= \langle \bar{s}s \rangle$.

The leading action in the odd-intrinsic-parity sector arises at ${\cal O}(p^4)$.
This is given by the chiral anomaly \cite{Wess:1971yu}
and explicitly stated by the Wess-Zumino-Witten ($WZW$) functional  ${\cal Z}_{WZW}[v,a]$ that can be
read in Ref.~\cite{Ecker:1994gg}. This contains all anomalous contributions to electromagnetic
and semileptonic meson decays. For completeness, we give the relevant terms to $\tau \to P \gamma \nu_\tau$ below
\bqa
\mathcal{L}_{WZW} = -\frac{i N_C}{48 \pi^2} \varepsilon_{\mu\nu\alpha\beta}
\langle \Sigma^\mu_L\, U^\dagger \,\partial^\nu r^\alpha\, U \,l^\beta +\Sigma^\mu_L \, l^\nu \,\partial^\alpha l^\beta
+\Sigma^\mu_L\, \partial^\nu l^\alpha\, l^\beta- (L \leftrightarrow R) \rangle\,,
\eqa
with $U=u^2$, $\Sigma^\mu_L= U^\dagger \partial^\mu U$ and $\Sigma^\mu_R = U \partial^\mu U^\dagger$.

It is well known \cite{Ecker:1988te,Cirigliano:2006hb} that higher orders in
the chiral expansion, i.e. even-intrinsic-parity ${\cal L}_{\chi PT}^{(n)}$
with $n>2$, bring in the information of heavier degrees of freedom that have
been integrated out, for instance resonance states. As our next step intends
to include the latter explicitly,
we will not consider higher orders in $\CPT$ in order to avoid double counting issues. In
order to fulfill this procedure ---at least, up to ${\cal O}(p^4)$ in the even-intrinsic-parity sector---
it is convenient to use the antisymmetric tensor representation for the
$J=1$ fields, as we comment below. Analogously, additional odd-intrinsic-parity amplitudes arise
at ${\cal O}(p^6)$ in $\CPT$, either from one-loop diagrams using
one vertex from the $WZW$ action or from tree-level
operators \cite{Bijnens:2001bb}. However we will assume that the latter are
fully generated by resonance contributions \cite{RuizFemenia:2003hm} and,
therefore, will not be included in the following.

The formulation of a Lagrangian theory that includes both the octet of
Goldstone mesons and $U(3)$ nonets of resonances is carried out through
the construction of a phenomenological Lagrangian \cite{CCWZ} where
chiral and discrete symmetries determine the
structure of the operators. Given the vector character of the Standard Model
(SM) couplings of the hadron matrix elements in $\tau$ decays, form factors
for these processes are ruled by vector and
axial-vector resonances. Notwithstanding those form factors are given,
in the $\tau \rightarrow P \gamma \nu_{\tau}$ decays, by a
three-point Green function where other quantum numbers might play a role,
namely scalar and pseudoscalar resonances \cite{Jamin:2000wn}.
However their
contribution should be very small for $\tau \rightarrow P \gamma \nu_{\tau}$.
This statement is based on the following observations: the scalar resonances will be irrelevant at tree level to the considered process due to
the discrete symmetry; about the pseudoscalar resonances, their contributions are suppressed due to first their heavy masses and also the fact that
the relevant couplings involving
the pseudoscalar resonances should be very tiny, since the decay of these states to $P \gamma$ has not been reported yet.
Thus in our description we include $J=1$ resonances only, and this is
done by considering a nonet of fields \cite{Ecker:1988te}~:
\begin{equation}
 R \,\equiv \, \frac{1}{\sqrt{2}} \, \sum_{i=0}^{8} \lambda_i \, \phi_{R,i}\; ,
\end{equation}
where $R = V, A$, stand for the vector and axial-vector resonance states. Under
the $SU(3)_L \otimes SU(3)_R$ chiral group, $R$ transforms as~:
\begin{equation} \label{eq:rtrans}
 R \, \rightarrow \, h(g,\varphi) \, R \, h(g,\varphi)^{\dagger} \; .
\end{equation}
The flavour structure of the resonances is analogous to that of the
Goldstone bosons in Eq.~(\ref{eq:phi_matrix}). One can also introduce the
covariant derivative
\begin{eqnarray}
\nabla_\mu X &  \equiv &     \partial_{\mu} X    + [\Gamma_{\mu}, X] \; \; , \\
\Gamma_\mu & = & \frac{1}{2} \, [ \,
u^\dagger (\partial_\mu - i r_{\mu}) u +
u (\partial_\mu - i \ell_{\mu}) u^\dagger \,] \; \; ,\nonumber
\end{eqnarray}
acting on any object $X$ that transforms as $R$ in Eq.~(\ref{eq:rtrans}),
like $u_{\mu}$ and $\chi_{\pm}$. The kinetic terms for the spin-one resonances
in the Lagrangian read \cite{Ecker:1988te}~:
\begin{equation} \label{eq:lag0}
{\cal L}_{ \rm kin}^R = -\frac{1}{2} \langle \,
\nabla^\lambda R_{\lambda\mu} \nabla_\nu R^{\nu\mu} \, \rangle + \frac{M_R^2}{4} \, \langle \,
R_{\mu\nu} R^{\mu\nu}\,  \rangle \; \; \; , \; \; \;  \; \; R \, = \, V,A \; ,
\end{equation}
$M_V$, $M_A$ being the masses of the nonets of vector and axial--vector
resonances in the chiral and large-$N_C$ limits, respectively.
Notice that we describe the resonance fields
through the antisymmetric tensor representation. With this description
one is able to collect, upon integration of the resonances, the bulk of the
low-energy couplings at ${\cal O}(p^4)$ in $\chi$PT without the inclusion of
additional local terms \cite{Ecker:1989yg}. So it is necessary
to use this representation if one does not include ${\cal L}_{\chi
PT}^{(4)}$ in the Lagrangian. Though analogous studies at higher
chiral orders have not been carried out, we will assume that no ${\cal
L}_{\chi PT}^{(n)}$ with $n=4,6,...$ in the even-intrinsic-parity and
$n=6,8,...$ in the odd-intrinsic-parity sectors need to be included in the
theory.

The construction of the interaction terms involving resonance and Goldstone fields is driven by
chiral and discrete symmetries with a generic structure given by~:
\begin{equation}
 {\cal O}_i \, \sim \, \langle \, R_1 R_2 ... R_j \, \chi^{(n)}(\varphi) \, \rangle \, ,
\end{equation}
where $\chi^{(n)}(\varphi)$ is a chiral tensor that includes only
Goldstone and auxiliary fields. It transforms like $R$ in
Eq.~(\ref{eq:rtrans}) and has chiral counting $n$ in the frame of $\CPT$.
This counting is relevant in the setting of the theory because, though the resonance
theory itself has no perturbative expansion, higher values of $n$ may originate
violations of the proper
asymptotic behaviour of form factors or Green functions. As a guide
we will include at least those operators that, contributing to our processes,
are leading when integrating out the resonances. In addition we do not include
operators with higher-order chiral tensors, $\chi^{(n)}(\varphi)$, that would
violate the $QCD$ asymptotic behaviour
unless their couplings are severely fine tuned to ensure the needed
cancellations of large momenta. In the odd-intrinsic-parity
sector, which contributes to the vector form factor, this amounts to include all
$\langle R \chi^{(4)} \rangle$ and $\langle RR \chi^{(2)} \rangle$ terms.
In the even-intrinsic-parity sector, contributing to the axial-vector form factors,
these are the terms $\langle R \chi^{(2)} \rangle$. However previous analyses
of the axial-vector contributions
\cite{GomezDumm:2003ku, Dumm:2009kj, Dumm:2009va, Cirigliano:2004ue} show the relevant role of the
$\langle RR \chi^{(2)} \rangle$ terms that, accordingly, are also considered
here.

We also assume exact $SU(3)$ symmetry in the construction of the interacting
terms, i.e. at the level of couplings. Deviations from exact symmetry in hadronic
tau decays have been considered in Ref.~\cite{Moussallam:2007qc}. However we do not
include $SU(3)$ breaking couplings because we are neither considering
next-to-leading order corrections in the $1/N_C$ expansion.

The lowest order interaction operators, linear in the resonance fields, have the structure
$\langle R \chi^{(2)}(\varphi)\rangle$. There are no odd-intrinsic-parity terms of this form.
The even-intrinsic-parity Lagrangian includes three coupling
constants \cite{Ecker:1988te}~:
\begin{eqnarray} \label{eq:lag1}
{\cal L}_2^{\mbox{\tiny V}} & = &  \frac{F_V}{2\sqrt{2}} \langle V_{\mu\nu}
f_+^{\mu\nu}\rangle + i\,\frac{G_V}{\sqrt{2}} \langle V_{\mu\nu} u^\mu
u^\nu\rangle  \; \, , \nonumber \\
{\cal L}_2^{\mbox{\tiny A}} & = &  \frac{F_A}{2\sqrt{2}} \langle A_{\mu\nu}
f_-^{\mu\nu}\rangle \;,
\end{eqnarray}
where
$f_\pm^{\mu\nu}  =  u F_L^{\mu\nu} u^\dagger \pm u^\dagger F_R^{\mu\nu}
u$ and $F_{R,L}^{\mu \nu}$ are the field strength tensors associated
with the external right- and left-handed auxiliary fields. All couplings
$F_V$, $G_V$ and $F_A$ are real.

The leading odd-intrinsic-parity operators, linear in the resonance fields, have
the form $\langle R \chi^{(4)}(\varphi)\rangle$. We will need those pieces that
generate the vertex with one vector resonance, a vector current and one pseudoscalar. The
minimal Lagrangian with these features is~:
\begin{equation}
\label{eq:l4odd}
{\cal L}_4^{\mbox{\tiny V}} = \sum_{i=1}^{7} \,\frac{c_i}{M_{V}}\,{\cal O}^i_{\mbox{\tiny{VJP}}} \, ,
\end{equation}
where $c_i$ are real dimensionless couplings, and the VJP operators
read \cite{RuizFemenia:2003hm}
\begin{eqnarray}
 {\cal O}_{\mbox{\tiny VJP}}^1 & = & \varepsilon_{\mu\nu\rho\sigma}\,
\langle \, \{V^{\mu\nu},f_{+}^{\rho\alpha}\} \nabla_{\alpha}u^{\sigma}\,\rangle
\; \; , \nonumber\\[2mm]
{\cal O}_{\mbox{\tiny VJP}}^2 & = & \varepsilon_{\mu\nu\rho\sigma}\,
\langle \, \{V^{\mu\alpha},f_{+}^{\rho\sigma}\} \nabla_{\alpha}u^{\nu}\,\rangle
\; \; , \nonumber\\[2mm]
{\cal O}_{\mbox{\tiny VJP}}^3 & = & i\,\varepsilon_{\mu\nu\rho\sigma}\,
\langle \, \{V^{\mu\nu},f_{+}^{\rho\sigma}\}\, \chi_{-}\,\rangle
\; \; , \nonumber\\[2mm]
{\cal O}_{\mbox{\tiny VJP}}^4 & = & i\,\varepsilon_{\mu\nu\rho\sigma}\,
\langle \, V^{\mu\nu}\,[\,f_{-}^{\rho\sigma}, \chi_{+}]\,\rangle
\; \; , \nonumber\\[2mm]
{\cal O}_{\mbox{\tiny VJP}}^5 & = & \varepsilon_{\mu\nu\rho\sigma}\,
\langle \, \{\nabla_{\alpha}V^{\mu\nu},f_{+}^{\rho\alpha}\} u^{\sigma}\,\rangle
\; \; ,\nonumber\\[2mm]
{\cal O}_{\mbox{\tiny VJP}}^6 & = & \varepsilon_{\mu\nu\rho\sigma}\,
\langle \, \{\nabla_{\alpha}V^{\mu\alpha},f_{+}^{\rho\sigma}\} u^{\nu}\,\rangle
\; \; , \nonumber\\[2mm]
{\cal O}_{\mbox{\tiny VJP}}^7 & = & \varepsilon_{\mu\nu\rho\sigma}\,
\langle \, \{\nabla^{\sigma}V^{\mu\nu},f_{+}^{\rho\alpha}\} u_{\alpha}\,\rangle
\;\; .
\label{eq:VJP}
\end{eqnarray}
Notice that we do not include analogous pieces with an axial-vector resonance, that
would contribute to the hadronization of the axial-vector current. This
has been thoroughly studied in Ref.~\cite{GomezDumm:2003ku} (see also Ref.~\cite{Dumm:2009va})
in the description of the $\tau \rightarrow \pi \pi \pi \nu_{\tau}$ process and it is shown
that no $\langle A \chi^{(4)}(\varphi) \rangle$ operators are needed to describe its hadronization.
Therefore those operators are not included in our minimal description of
the relevant form factors.

In order to study tau decay processes with a pseudoscalar meson and a photon in the
final state one also has to consider non-linear terms in the resonance
fields. Indeed the hadron final state in $\tau \rightarrow P \gamma
\nu_{\tau}$ decays can be driven by vertices involving two resonances and a
pseudoscalar meson. The structure of
the operators that give those vertices is $\langle R_1 R_2
\chi^{(2)}(\varphi) \rangle$, and has been worked out before
\cite{GomezDumm:2003ku, RuizFemenia:2003hm}. They include both even- and
odd-intrinsic-parity terms~:
\begin{equation}
\label{eq:lag21}
{\cal L}_2^{\mbox{\tiny RR}} \, = \, \sum_{i=1}^{5} \, \lambda_i \,
{\cal O}^i_{\mbox{\tiny VAP}} \;
\; + \;
\sum_{i=1}^{4} \,d_i\,{\cal O}^i_{\mbox{\tiny{VVP}}}\; ,
\end{equation}
where $\lambda_i$, and $d_i$ are unknown real dimensionless couplings.
The operators ${\cal O}^i_{\mbox{\tiny RRP}}$ are given by~:
\begin{itemize}
\item[1/] VAP terms
\begin{eqnarray}
\label{eq:VAP}
{\cal O}^1_{\mbox{\tiny VAP}} &  = & \langle \,  [ \, V^{\mu\nu} \, , \,
A_{\mu\nu} \, ] \,  \chi_- \, \rangle \; \; , \nonumber \\ [2mm]
{\cal O}^2_{\mbox{\tiny VAP}} & = & i\,\langle \, [ \, V^{\mu\nu} \, , \,
A_{\nu\alpha} \, ] \, h_\mu^{\;\alpha} \, \rangle \; \; , \\ [2mm]
{\cal O}^3_{\mbox{\tiny VAP}} & = &  i \,\langle \, [ \, \nabla^\mu V_{\mu\nu} \, , \,
A^{\nu\alpha}\, ] \, u_\alpha \, \rangle \; \; ,  \nonumber \\ [2mm]
{\cal O}^4_{\mbox{\tiny VAP}} & = & i\,\langle \, [ \, \nabla^\alpha V_{\mu\nu} \, , \,
A_\alpha^{\;\nu} \, ] \,  u^\mu \, \rangle \; \; , \nonumber \\ [2mm]
{\cal O}^5_{\mbox{\tiny VAP}} & =  & i \,\langle \, [ \, \nabla^\alpha V_{\mu\nu} \, , \,
A^{\mu\nu} \, ] \, u_\alpha \, \rangle \nonumber \; \; .
\end{eqnarray}
with $h_{\mu \nu} = \nabla_{\mu} u_{\nu} + \nabla_{\nu} u_{\mu}$, and
\item[2/] VVP terms
\begin{eqnarray}
{\cal O}_{\mbox{\tiny VVP}}^1 & = & \varepsilon_{\mu\nu\rho\sigma}\,
\langle \, \{V^{\mu\nu},V^{\rho\alpha}\} \nabla_{\alpha}u^{\sigma}\,\rangle
\; \; , \nonumber\\[2mm]
{\cal O}_{\mbox{\tiny VVP}}^2 & = & i\,\varepsilon_{\mu\nu\rho\sigma}\,
\langle \, \{V^{\mu\nu},V^{\rho\sigma}\}\, \chi_{-}\,\rangle
\; \; , \nonumber\\[2mm]
{\cal O}_{\mbox{\tiny VVP}}^3 & = & \varepsilon_{\mu\nu\rho\sigma}\,
\langle \, \{\nabla_{\alpha}V^{\mu\nu},V^{\rho\alpha}\} u^{\sigma}\,\rangle
\; \; , \nonumber\\[2mm]
{\cal O}_{\mbox{\tiny VVP}}^4 & = & \varepsilon_{\mu\nu\rho\sigma}\,
\langle \, \{\nabla^{\sigma}V^{\mu\nu},V^{\rho\alpha}\} u_{\alpha}\,\rangle
\; \; .
\label{eq:VVP}
\end{eqnarray}
\end{itemize}
We emphasize that ${\cal L}_2^{\mbox{\tiny RR}}$  is a complete basis for constructing vertices with only
one pseudoscalar meson; for a larger number of pseudoscalars additional operators
might be added. As we are only interested in tree-level diagrams, the equation of
motion arising from ${\cal L}_{\chi PT}^{(2)}$ in Eq.~(\ref{eq:op2}) has been used in
${\cal L}_4^{\mbox{\tiny V}}$ and ${\cal L}^{\mbox{\tiny RR}}_2$ to eliminate superfluous operators.
\par
Hence our theory is given by the Lagrangian~:
\begin{equation}
\label{eq:ourtheory}
 {\cal L}_{R \chi T} \, = \, {\cal L}_{\chi PT}^{(2)} \,+{\cal L}_{WZW} \,  + \, {\cal L }_{\mbox{\tiny kin}}^{\mbox{\tiny R}} \, + \,
{\cal L}_2^{\mbox{\tiny A}} \, + \,{\cal L}_2^{\mbox{\tiny V}} \, + \,
 {\cal L}_4^{\mbox{\tiny V}} \, + \, {\cal L}_2^{\mbox{\tiny RR}} \, .
\end{equation}
It is important to point out that the resonance theory constructed above is
not a theory of $QCD$ for arbitrary values of the couplings in the interaction
terms. As we will see later on, these constants can be constrained by
imposing well accepted dynamical properties of the underlying theory.

\section{Matrix element decomposition, kinematics and decay rate}\label{Kinematics}

\begin{figure}[h!]
\centering
\includegraphics[scale=0.5]{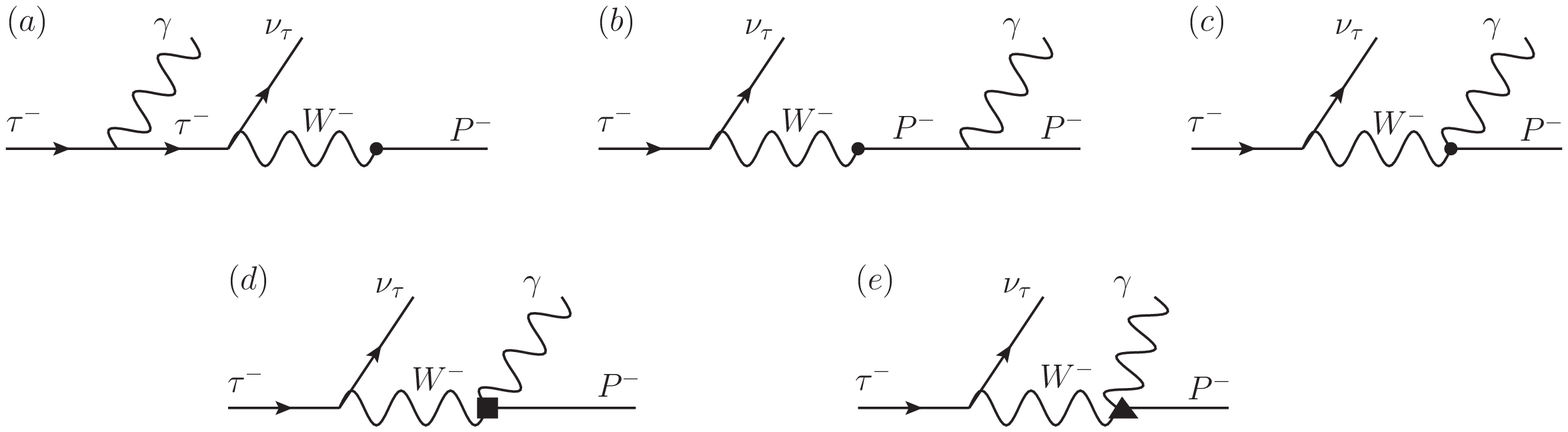}
\caption{Feynman diagrams for the different kinds of contributions to the radiative decays of the tau including one meson, as explained in the main text. The
 dot indicates the hadronization of the $QCD$ currents. The solid square represents the $SD$ contribution mediated by the vector current and the solid
 triangle the $SD$ contribution via the axial-vector current.}
\label{diagrams general decomposition amplitude in radiative decays tau with one meson}
\end{figure}

The process we are going to compute is $\tau^-(p_\t) \to \nu_\tau(q) P^-(p) \gamma(k)\,$. The kinematics of this decay is equivalent to that of the radiative pion decay
 \cite{Brown:1964zza}. We will use $t := (p_\t - q)^2 = (k + p)^2$. In complete analogy to the case of the radiative pion decay \cite{Bae67}, the matrix element
 for the decay of $\tau^- \to P^- \gamma \nu_\t$ can be written as the sum of four contributions:
\begin{equation}
   \mathcal{M} \left[\tau^-(p_\t) \to \nu_\tau(q) P^-(p) \gamma(k)\right]
   = \mathcal{M}_{IB_\tau} + \mathcal{M}_{IB_P} + \mathcal{M}_{V} + \mathcal{M}_{A}\,,
\end{equation}
with \footnote{Notice that $i$ and minus factors differ with respect to Ref.~\cite{Decker:1993ut} ($DF$). Moreover, our form factors have dimension of inverse mass
while theirs are dimensionless. In their work, the factor of $(\sqrt{2}m_\pi)^{-1}$ in the form factors is compensated by defining the sum over polarizations
of the matrix element squared with an extra $2 m_\pi^2$ factor. This should be taken into account to compare formulae in both works using that
$F_V(t)^{DF}\,=\,\sqrt{2}m_\pi F_V(t)^{Our}\,,\;\;F_A(t)^{DF}\,=\,2\sqrt{2}m_\pi F_A(t)^{Our}$.}
\begin{eqnarray} \label{Gral_IB}
  i \mathcal{M}_{IB_{\tau}} & = &  G_F\, V_{CKM}^{ij}\, e\, F_P\, p_\mu\,
\epsilon_\nu(k)\, L^{\mu \nu}\,,
   \nonumber \\
  i \mathcal{M}_{IB_{P}} & = & \, G_F\, V_{CKM}^{ij}\, e\, F_P \,\,\epsilon^\nu(k)\,
\left( \frac{2\,p_\nu \, (k\, +\, p)_\mu}
      {m_P^2\, -\, t} \,+\, g_{\mu\nu} \right)\, L^\mu\,, \nonumber \\
  i \mathcal{M}_V & = &  i \,G_F \,V_{CKM}^{ij}\, e\, F_V(t)\, \varepsilon_{\mu \nu \rho \sigma}
     \epsilon^\nu(k)\,  k^\rho\, p^\sigma\,  L^\mu\,,  \nonumber \\
  i \mathcal{M}_{A} & = & G_F\, V_{CKM}^{ij}\, e\, F_A(t)\, \epsilon^\nu(k)\, \left[(t-m_P^2) \,g_{\mu\nu}-2p_\nu\, k_\mu \, \right] L^\mu\,,
\end{eqnarray}
where $\epsilon_\nu$ is the polarization vector of the photon. $F_V(t)$ and $F_A(t)$ are the so-called structure dependent form
factors. Finally $L^\mu$ and $L^{\mu\nu}$ are lepton currents defined by
\begin{eqnarray}
   L^\mu & = & \bar{u}_{\nu_\t}(q) \gamma^\mu (1-\gamma_5) u_\t(p_\t)\,, \nonumber \\
   L^{\mu \nu} & = & \bar{u}_{\nu_\t}(q) \gamma^\mu (1-\gamma_5)
\frac{\strich{k} - \strich{p}_\t
      - M_\t}{(k - p_\t)^2 - M_\t^2} \gamma^\nu u_\t(p_\t)\,.
\end{eqnarray}
The notation introduced for the amplitudes describes the four kinds of contributions: $\mathcal{M}_{IB_{\tau}}$ is the bremsstrahlung off
the tau, (Figure \ref{diagrams general decomposition amplitude in radiative decays tau with one meson}(a)); $\mathcal{M}_{IB_{P}}$ is the sum of
the bremsstrahlung off the $P$-meson (Figure \ref{diagrams general decomposition amplitude in radiative decays tau with one meson}(b)), and the
seagull diagram (Figure \ref{diagrams general decomposition amplitude in radiative decays tau with one meson}(c)); $\mathcal{M}_{V}$ is the
structure dependent vector contribution (Figure \ref{diagrams general decomposition amplitude in radiative decays tau with one meson}(d)) and $\mathcal{M}_{A}$
the structure dependent axial-vector contribution (Figure \ref{diagrams general decomposition amplitude in radiative decays tau with one meson}(e)).
Our ignorance of the exact mechanism of hadronization is parametrized in terms of the two form factors $F_A(t)$ and $F_V$(t). In fact, these form
factors are the same functions of the momentum transfer $t$ as those in the radiative pion decay, the only difference being that $t$ now varies
from $0$ up to $M_\tau^2$ rather than just up to $m_\pi^2$.

The two matrix elements $\mathcal{M}_{IB_{\tau}}$ and $\mathcal{M}_{IB_{P}}$ are not separately gauge invariant, but their sum, ie. the (total)
matrix element for internal bremsstrahlung $IB$
\begin{equation}
   \mathcal{M}_{IB} = \mathcal{M}_{IB_{\tau}} + \mathcal{M}_{IB_{P}}\,,
\end{equation}
is indeed gauge invariant, as $\mathcal{M}_V$ and $\mathcal{M}_A$ are. We also define the (total) structure dependent radiation $SD$ by
\begin{equation}
   \mathcal{M}_{SD} = \mathcal{M}_{V} + \mathcal{M}_{A}\,.
\end{equation}
The spinor structure can be rearranged to give
\begin{eqnarray}
  i \mathcal{M}_{IB} & = & G_F\, V_{ij}^{CKM}\, e\, F_P\, M_\t \bar{u}_{\nu_\t}(q) (1+\gamma_5) \left[\frac{p_\t \cdot \epsilon}{p_\t \cdot k} - \frac{p \cdot \epsilon}
{p \cdot k} - \frac{\strich{k}\strich{\epsilon}}{2 p_\t \cdot k} \right] u_\t(p_\t)\,,  \\
  i \mathcal{M}_{SD} & = & G_F\, V_{ij}^{CKM}\, e \left\{ i \varepsilon_{\mu \nu \rho \sigma} L^\mu \epsilon^\nu k^\rho p^\sigma F_V(t)
      + \bar{u}_{\nu_\t}(q) (1+\gamma_5) \left[ (t-m_P^2) \strich{\epsilon} - 2(\epsilon \cdot p) \strich{k} \right] u(p_\t) F_A(t) \right\} \,. \nonumber
\end{eqnarray}
The square of the matrix element is then given by
\begin{equation}
  \overline{| \mathcal{M} |^2} = \overline{| \mathcal{M}_{IB} |^2} + 2 \overline{\Re e (\mathcal{M}_{IB} \mathcal{M}_{SD}^\star)} + \overline{| \mathcal{M}_{SD} |^2}\,,
\end{equation}
where the bar denotes summing over the photon polarization, the neutrino spin and averaging over the tau spin.

We follow Ref.~\cite{Decker:1993ut} and divide the decay rate as follows: the internal bremsstrahlung part $\Gamma_{IB}$ arising from
$\overline{| \mathcal{M}_{IB} |^2}$, the structure dependent part $\Gamma_{SD}$ coming from $\overline{| \mathcal{M}_{SD} |^2}$, and the interference part
$\Gamma_{INT}$ stemming from $2 \overline{\Re e (\mathcal{M}_{IB} \mathcal{M}_{SD}^\star)}$. Furthermore $\Gamma_{SD}$ is subdivided into the vector-vector
($\Gamma_{VV}$), the axial-vector--axial-vector ($\Gamma_{AA}$) and the vector--axial-vector interference term ($\Gamma_{VA}$). Similarly $\Gamma_{INT}$ gets
split into the internal bremsstrahlung-vector interference ($\Gamma_{IB-V}$) and the internal bremsstrahlung--axial-vector interference ($\Gamma_{IB-A}$) parts.
 Thus, one has
\begin{eqnarray} \label{parts Gamma radiative decay tau one pG}
    \Gamma_{ALL} & = & \Gamma_{IB} + \Gamma_{SD} + \Gamma_{INT}\,, \nonumber \\
    \Gamma_{SD} & = & \Gamma_{VV} + \Gamma_{VA} + \Gamma_{AA}\,, \nonumber \\
    \Gamma_{INT} & = & \Gamma_{IB-V} + \Gamma_{IB-A}\,.
\end{eqnarray}
It is convenient to use the dimensionless variables $x$ and $y$ to proceed, as used in Ref.\cite{Decker:1993ut} and references therein:
\begin{equation}
 x := \frac{2 p_\tau \cdot k}{M_\t^2}\,,\quad \quad y := \frac{2 p_\tau \cdot p}{M_\t^2}\,.
\end{equation}
In the tau rest frame $x$ and $y$ are the energies $E_\gamma$ and $E_\pi$ of the photon and the pion, respectively, expressed in units of $M_\t/2$:
\begin{equation} \label{x,y_defs}
   E_\gamma = \frac{M_\t}{2} x \,,\quad \quad E_\pi    = \frac{M_\t}{2} y\,.
\end{equation}
 Eq. (\ref{x,y_defs}) sets the scale for the photons to be considered as ''hard'' or ''soft''. This means that the formulae for internal bremsstrahlung should be
similar for radiative tau and pion decay, once they are expressed in terms of $x$ and $y$, as it is the case, albeit photons of comparable softness will
have very different energies in both cases.

The kinematical boundaries for $x$ and $y$ are given by
\begin{eqnarray}
   0 \leq x \leq & 1 - r_P^2\,,\quad \quad 1 - x + \frac{r_P^2}{1 -x} \leq y \leq 1 + r_P^2\,,
\end{eqnarray}
where
\begin{equation}
 r_P^2:= \left( \frac{m_P}{M_\t} \right)^2 \sim ^{0\mathrm{.}006}_{0\mathrm{.}077} \ll 1\,,
\end{equation}
where the upper figure corresponds to $P=\pi$ and the lower one to $P=K$. It is also useful to note that
\begin{equation}
   p \cdot k = \frac{M_\t^2}{2} (x + y - 1 - r_P^2)\,,\quad\quad   t := (p_\t - q)^2 = (k + p)^2 = M_\t^2 (x + y - 1)\,.
\end{equation}
The differential decay rate is given by \cite{PartKin}
\begin{equation}
  d\Gamma(\tau^- \rightarrow \nu_\tau P^- \gamma) =
  \frac{1}{512 \pi^5 E_{\tau}} \delta^{(4)} (k + p + q - p_\t)
  \overline{| \mathcal{M} |^2} \frac{d^3 \vec{k} d^3 \vec{p} d^3 \vec{q}}{E_{\gamma} E_{\pi} E_{\nu}}\,,
\end{equation}
where the bar over the matrix element denotes summing over the photon polarization, the neutrino spin and averaging over the tau spin. Choice of the tau rest frame,
integration over the neutrino momentum, $\vec{p}$, and the remaining angles and introduction of $x$ and $y$ yield
\begin{equation}
   \frac{d^2 \Gamma}{dx\, dy} = \frac{m_\t}{256 \pi^3} \overline{| \mathcal{M} |^2}\,.
\end{equation}
The integration over $y$ yields the photon spectrum
\begin{equation}
    \frac{\mathrm{d}\Gamma}{\mathrm{d}x} = \int_{1 - x + \frac{r_P^2}{1-x}}^{1 + r_P^2} \mathrm{d}y \, \frac{\mathrm{d}^2 \Gamma}{\mathrm{d}x\, \mathrm{d}y}\,.
\end{equation}
A low-energy cut must be introduced for the photon energy because of the infrared divergence of the internal bremsstrahlung. By requiring $x \geq x_0$ one
obtains the integrated decay rate
\begin{equation}
   \Gamma (x_0) = \Gamma (E_0) = \int_{x_{0}}^{1 - r_P^{2}} \mathrm{d}x \frac{\mathrm{d} \Gamma}{\mathrm{d}x}\,,
\end{equation}
that does depend on the photon energy cut-off ($E_0 = \frac{M_\t}{2} x_0$). Instead of $x$ and $y$ one can also use $x$ and $z$, where $z$ is the scaled momentum
 transfer squared:
\begin{equation}
   z = \frac{t}{M_\t^2} = x + y - 1\,,
\end{equation}
whose kinematical boundaries are
\begin{equation}
   z - r_P^2 \leq x  \leq 1 - \frac{r_P^2}{z}\,,\quad \quad r_P^2 \leq z \leq 1\,.
\end{equation}
Integration of $\frac{\mathrm{d}^2 \Gamma}{\mathrm{d}x\, \mathrm{d}y}$ over $x$ yields the spectrum in $z$, i. e. the spectrum in the invariant mass of
 the meson-photon system:
\begin{equation}
\frac{\mathrm{d}\Gamma}{dz} (z) = \frac{\mathrm{d}\Gamma}{\mathrm{d}z} \left(\sqrt{t}\right) = \int_{z - r_P^2}^{1 - r_P^2/z}
\mathrm{d}x\frac{\mathrm{d}^2 \Gamma}{\mathrm{d}x\, \mathrm{d}y} (x, y=z-x+1)\,.
\end{equation}
The integrated rate for events with $t \geq t_0$ is then given by
\begin{equation}
   \Gamma(z_0) = \Gamma\left(\sqrt{t_0}\right) = \int_{z_0}^{1} \mathrm{d}z \frac{\mathrm{d}\Gamma}{\mathrm{d}z} (z)\,.
\end{equation}
We note that $z_0$ is both an infrared and a collinear cut-off.

In terms of the quantities defined in Eq.~(\ref{parts Gamma radiative decay tau one pG}) the differential decay rate is
\begin{eqnarray}
   \frac{\mathrm{d}^2\Gamma_{IB}}{\mathrm{d}x \, \mathrm{d}y} & = & \frac{\alpha}{2 \pi} f_{IB}\left(x,y,r_P^2\right) \frac{\gnr}{\left(1-r_P^2\right)^2}\,, \\
   \frac{\mathrm{d}^2\Gamma_{SD}}{\mathrm{d}x \, \mathrm{d}y} & = & \frac{\alpha}{8 \pi} \frac{M_\t^4}{F_P^2} \left[ |F_V(t)|^2 f_{VV}\left(x,y,r_P^2\right) +
 4 \Re e(F_V(t) F_A^\star(t))f_{VA}\left(x,y,r_P^2\right) +\right.\nonumber\\
& & \left. 4 |F_A(t)|^2 f_{AA}(x,y,r_P^2) \right] \frac{\gnr}{(1-r_P^2)^2}\,, \nonumber \\
   \frac{\mathrm{d}^2\Gamma_{INT}}{\mathrm{d}x \, \mathrm{d}y} & = &\frac{\alpha}{2 \pi}\frac{M_\t^2}{F_P}\left[ f_{IB-V}\left(x,y,r_P^2\right) \Re e(F_V(t))
 +2 f_{IB-A}\left(x,y,r_P^2\right) \Re e(F_A(t)) \right] \frac{\gnr}{\left(1-r_P^2\right)^2}\,,\nonumber
\end{eqnarray}
with $\alpha = e^2/(4\pi)$ standing for the fine structure constant and
\begin{eqnarray}\label{kinematicsF}
   f_{IB} \left(x,y,r_P^2\right) & = & \frac{[r_P^4 (x + 2) - 2 r_P^2 (x + y) + (x + y - 1)\left(2 - 3x + x^2 + xy\right)]\left(r_P^2 - y + 1\right)}
   {\left(r_P^2 - x - y +1\right)^2 x^2} \,,\nonumber \\
   f_{VV} \left(x,y,r_P^2\right) & = & - [r_P^4 (x + y) + 2 r_P^2 (1 - y) (x + y) + (x + y - 1)\left(-x + x^2 - y + y^2\right)]\,,\nonumber \\
   f_{AA} \left(x,y,r_P^2\right) & = & f_{VV}\left(x,y,r_P^2\right)\,,\nonumber \\
   f_{VA}\left(x,y,r_P^2\right) & = & - [r_P^2 (x + y) + (1 - x - y)(y-x)] \left(r_P^2 - x - y + 1\right) \,, \nonumber \\
   f_{IB-V}\left(x,y,r_P^2\right) & = & - \frac{\left(r_P^2 - x - y + 1\right)\left(r_P^2 - y + 1\right)}{x} \,,\nonumber \\
   f_{IB-A}\left(x,y,r_P^2\right) & = & -\frac{[r_P^4 - 2 r_P^2(x + y) + (1 - x + y) (x + y - 1)]\left(r_P^2 - y + 1\right)}{\left(r_P^2 - x - y + 1\right) x}\,.
\end{eqnarray}
The radiative decay rate has been expressed in term of the width of the non-radiative decay ($\tau^- \to \nu_\tau P^-$):
\begin{equation} \label{definition Gamma non radiative}
   \gnr = \frac{G_F^2  |V_{CKM}^{ij}|^2 F_P^2}{8 \pi} M_\t^3 (1 - r_P^2)^2\,.
\end{equation}
We finish this section by presenting the analytical expressions for the invariant mass spectrum:
\begin{eqnarray}\label{kinematicsZ}
   \frac{\mathrm{d}\Gamma_{IB}}{\mathrm{d}z} & = & \frac{\alpha}{2 \pi} \left[ r_P^4 (1 -z) + 2 r_P^2 \left(z - z^2\right)
   - 4z + 5 z^2 - z^3 \right.\nonumber \\
   &  & \left. + \left(r_P^4 z + 2 r_P^2 z - 2z -2z^2 + z^3\right) \mathrm{ln} z \right] \frac{1}{z^2 - r_P^2 z} \frac{\gnr}{\left(1 - r_P^2\right)^2}\,,\nonumber \\
   \frac{\mathrm{d}\Gamma_{VV}}{\mathrm{d}z} & = & \frac{\alpha}{24 \pi} \frac{M_\t^4}{F_P^2} \frac{(z - 1)^2 \left(z - r_P^2\right)^3 (1 + 2z)}{z^2} |F_V(t)|^2
\frac{\gnr}{(1 - r_P^2)^2} \,,\nonumber \\
   \frac{\mathrm{d}\Gamma_{VA}}{\mathrm{d}z} & = & 0 \,,\nonumber \\
   \frac{\mathrm{d}\Gamma_{AA}}{\mathrm{d}z} & = &  \frac{\alpha}{6 \pi} \frac{M_\t^4}{F_P^2} \frac{(z - 1)^2 \left(z - r_P^2\right)^3 (1 + 2z)}{z^2} |F_A(t)|^2
   \frac{\gnr}{\left(1 - r_P^2\right)^2}\,, \nonumber \\
   \frac{\mathrm{d}\Gamma_{IB-V}}{\mathrm{d}z} & = & \frac{\alpha}{2 \pi} \frac{M_\t^2}{F_P} \frac{(z-r_P^2)^2 (1 -z  + z \ln z)}{z} \Re e(F_V(t))
   \frac{\gnr}{\left(1 - r_P^2\right)^2}\,, \nonumber \\
   \frac{\mathrm{d}\Gamma_{IB-A}}{\mathrm{d}z} & = & -\frac{\alpha}{\pi} \frac{M_\t^2}{F_P} \left[ r_P^2 (1 - z) - 1 -z + 2 z^2  \right. \nonumber \\
   & & \left. + \left(r_P^2 z - 2z - z^2\right) \ln z \right]\frac{z-r_P^2}{z} \Re e(F_A(t)) \frac{\gnr}{\left(1 - r_P^2\right)^2}\,.
\end{eqnarray}
The interference terms $IB-V$ and $IB-A$ are now finite in the limit $z \to r_P^2$, which proves that their infrared divergences are integrable.

Although the above formulae have been noted in Ref.\cite{Decker:1993ut}, we independently calculate them
\footnote{ We disagree with Ref.~\cite{Decker:1993ut} on the signs of $f_{VA}$ and $f_{IB-V}$ in Eq.(\ref{kinematicsF}) and $\frac{\mathrm{d}\Gamma_{IB-V}}{\mathrm{d}z}$
in Eq.(\ref{kinematicsZ}), even after taking into account the minus sign difference in the definition of the $IB$ part. }
and explicitly give them here for completeness. Moreover we would like to point out that due to the
fact that our definitions of the form-factors $F_V(t)$ and $F_A(t)$ differ from the ones given in Ref.\cite{Decker:1993ut},
as we have mentioned before, there are some subtle differences in the above formulae between ours and theirs.

\section{Structure dependent form factors in $\tau^- \to \pi^- \gamma \nu_\tau$} \label{SDFFpi}

\begin{figure}[ht]
\begin{center}
\includegraphics[scale=0.7]{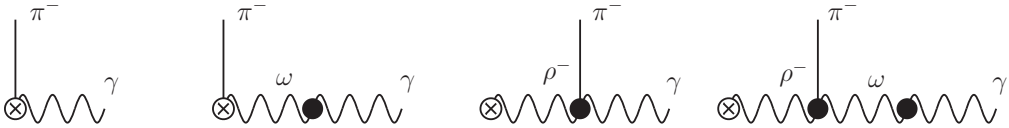}
\caption{Vector current contributions to $\tau^-\rightarrow  \pi^- \gamma \nu_\tau$. \label{fig.pi.v}}
\end{center}
\end{figure}

The Feynman diagrams, which are relevant to the vector current contributions to the $SD$ part of the
$\tau^- \to \pi^- \gamma \nu_\tau$ processes are given in Figure \ref{fig.pi.v}. The analytical result is found to be
\begin{equation}
i \mathcal{M}_{\mathrm{SD}_{V}}= i G_F\, V_{ud}\, e \,\overline{u}_{\nu_\tau}(q)\,\gamma^\mu (1-\gamma_5)\,u_\tau(s)
\varepsilon_{\mu\nu\alpha\beta} \,\epsilon^{\nu}(k)\, k^\alpha p^\beta\, F_V^\pi(t)\,,
\end{equation}
where the vector form-factor $F_V^\pi(t)$ is
\begin{eqnarray}
F_V^\pi(t) &=& -\frac{N_C}{24\pi^2 F_\pi}+ \frac{2\sqrt2 F_V}{3 F_\pi M_V
}\bigg[ (c_2-c_1-c_5) t +
(c_5-c_1-c_2-8c_3) m_\pi^2 \bigg]\times\nonumber \\
& &  \left[ \frac{\mathrm{cos}^2\theta}{M_\phi^2}\left(1-\sqrt{2} \mathrm{tg}\theta \right)
+ \frac{\mathrm{sin}^2\theta}{M_\omega^2}\left(1+\sqrt{2} \mathrm{cotg}\theta \right)\right]
\nonumber \\
& & + \frac{2\sqrt2 F_V}{3 F_\pi M_V }\, D_\rho(t)\,  \bigg[ ( c_1-c_2-c_5+2c_6) t +
(c_5-c_1-c_2-8c_3) m_\pi^2 \bigg] \nonumber \\
& & + \frac{4 F_V^2}{3 F_\pi }\, D_\rho(t)\,  \bigg[ d_3 t +
(d_1+8d_2-d_3) m_\pi^2 \bigg]\times\nonumber \\
& & \left[ \frac{\mathrm{cos}^2\theta}{M_\phi^2}\left(1-\sqrt{2} \mathrm{tg}\theta \right)
+ \frac{\mathrm{sin}^2\theta}{M_\omega^2}\left(1+\sqrt{2} \mathrm{cotg}\theta \right)\right]\,.
\nonumber \\
\end{eqnarray}
Here we have defined 
$D_R(t)$ as
\begin{equation}
D_R(t) = \frac{1}{M_R^2 - t - i M_R \Gamma_R(t)}\,.
\end{equation}
$\Gamma_R(t)$ stands for the decay width of the resonance $R$, which will be included following Refs. \cite{Dumm:2009va, GomezDumm:2000fz}. For completeness,
we write the explicit expressions of the off-shell widths in Appendix \ref{appendix-width}.

We will assume the ideal mixing case for the vector resonances $\omega$ and $\phi$ in any numerical application:
\begin{eqnarray}
\omega_1 = \mathrm{cos}\theta \;\omega - \mathrm{sin}\theta\;\phi \; \sim \sqrt{\frac{2}{3}} \omega - \sqrt{\frac{1}{3}} \phi \,, \nonumber \\
\omega_8 = \mathrm{sin}\theta \;\omega + \mathrm{cos}\theta\;\phi \; \sim \sqrt{\frac{2}{3}} \phi + \sqrt{\frac{1}{3}} \omega \,.
\end{eqnarray}
\\

\begin{figure}[ht]
\begin{center}
\includegraphics[scale=0.65]{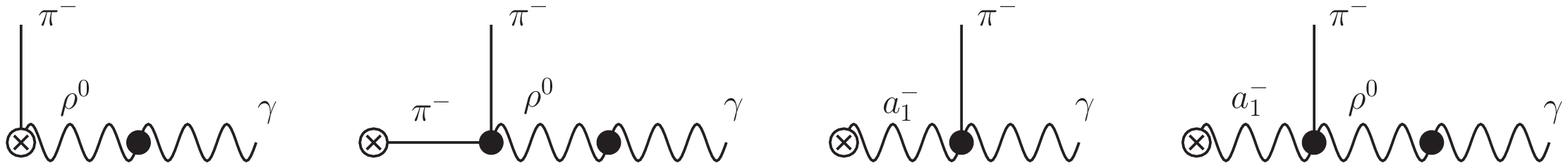}
\caption{Axial-vector current contributions to $\tau^-\rightarrow  \pi^- \gamma \nu_\tau$. \label{fig.pi.a}}
\end{center}
\end{figure}

The Feynman diagrams related to the axial-vector current contribution to the $SD$ part are given in Figure \ref{fig.pi.a}.
The corresponding result is
\begin{equation}
i \mathcal{M}_{\mathrm{SD}_{A}} =  G_F\, V_{ud}\, e \,\overline{u}_{\nu_\tau}(q)\,\gamma^\mu (1-\gamma_5)\,u_\tau(s)
 \,\epsilon^{\nu}(k)\, \big[ (t-m_\pi^2)g_{\mu\nu}-2k_\mu p_\nu \big]\, F_A^\pi(t)\,,
\end{equation}
where the axial-vector form-factor $F_A^\pi(t)$ is
\begin{eqnarray} \label{fapit}
F_A^\pi(t) &=& \frac{F_V^2}{2F_\pi M_\rho^2}\left(1-\frac{2G_V}{F_V}\right) - \frac{F_A^2}{ 2 F_\pi} D_{\mathrm{a}_1}(t)
+ \frac{\sqrt2 F_A F_V}{ F_\pi M_\rho^2 }\, D_{\mathrm{a}_1}(t)\,  \bigg( - \lambda'' t +
\lambda_0 m_\pi^2 \bigg)\,,\nonumber\\
\end{eqnarray}
where we have used the notation
\begin{eqnarray}
\sqrt{2}\lambda_0 &=&-4\lambda_1- \lambda_2-\frac{\lambda_4}{2}-\lambda_5\,, \nonumber \\
\sqrt{2} \lambda''  &=& \lambda_2-\frac{\lambda_4}{2}-\lambda_5\,,
\end{eqnarray}
for the relevant combinations of the couplings in $\mathcal{L}_2^{VAP}$, Eq. (\ref{eq:VAP}).

\section{Structure dependent form factors in  $\tau^- \to K^- \gamma \nu_\tau$} \label{SDFFK}

Although one can read the following observation from Eq.(\ref{Gral_IB}), let us emphasize that the model independent part
$\mathcal{M}_{\mathrm{IB}_{\tau+K}}$ is the same as in the pion case by replacing the pion decay constant $F_\pi$ with the kaon decay constant
 $F_K$.  A brief explanation about this replacement is in order. The difference of  $F_\pi$ and $F_K$ is generated by the low-energy constants
 and the chiral loops in $\CPT$~\cite{Gasser:1984gg}, while in the large $N_C$ limit of $\RCT$ this difference is due to
 the scalar resonances in an implicit way. Due to the scalar tadpole, one can always attach a scalar resonance to any of the pseudo-Goldstone
boson fields, which will cause its wave function renormalization. A convenient way to count this effect is to make the scalar field redefinition
before the explicit computation to eliminate the scalar tadpole effects. In the latter method, one can easily get the difference of $F_\pi$ and
$F_K$. For details, see Ref.~\cite{SanzCillero:2004sk} and references therein. For the model dependent parts, the simple replacements are not
applicable and one needs to work out the corresponding form factors explicitly.

\begin{figure}[ht]
\begin{center}
\includegraphics[scale=0.7]{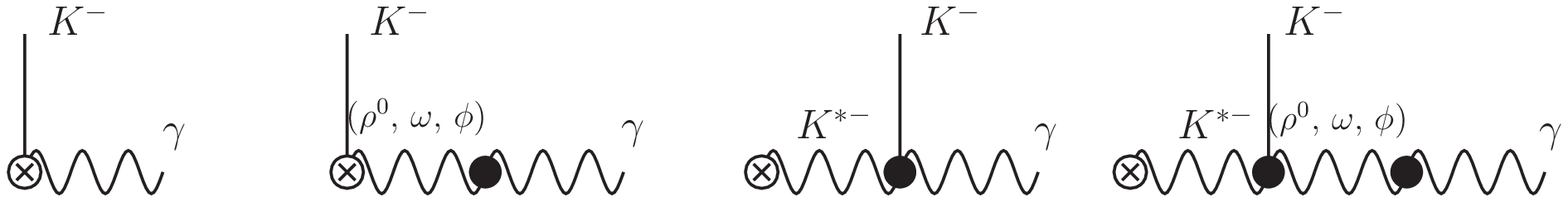}
\caption{Vector current contributions to $\tau^-\rightarrow  K^- \gamma \nu_\tau$. \label{fig.k.v}}
\end{center}
\end{figure}

The vector current contributions to the $SD$ part of the $\tau^- \to K^- \gamma \nu_\tau$ process are given in
Figure \ref{fig.k.v}. The analytical result is found to be
\begin{equation}
i \mathcal{M}_{\mathrm{SD}_{V}}= i G_F\, V_{us}\, e \,\overline{u}_{\nu_\tau}(q)\,\gamma^\mu (1-\gamma_5)\,u_\tau(s)
\varepsilon_{\mu\nu\alpha\beta} \,\epsilon^{\nu}(k)\, k^\alpha p^\beta\, F_V^K(t)\,,
\end{equation}
where the vector form-factor $F_V^K(t)$ is
\begin{eqnarray}
F_V^K(t) &=& -\frac{N_C}{24\pi^2 F_K}+ \frac{\sqrt2 F_V}{ F_K M_V }\bigg[ (c_2-c_1-c_5) t +
(c_5-c_1-c_2-8c_3) m_K^2 \bigg]\times\nonumber\\
& & \left[ \frac{1}{M_\rho^2}-\frac{\mathrm{sin}^2\theta}{3M_\omega^2}
\left( 1-2\sqrt{2} \mathrm{cotg}\theta\right) -\frac{\mathrm{cos}^2\theta}{3M_\phi^2} \left( 1+2\sqrt{2} \mathrm{tg}\theta\right)\right]
\nonumber \\ & &
+ \frac{2\sqrt2 F_V}{3 F_K M_V }\, D_{K^*}(t)\,  \bigg[ ( c_1-c_2-c_5+2c_6) t +
(c_5-c_1-c_2-8c_3) m_K^2
\nonumber \\ & &
+24 c_4(m_K^2-m_\pi^2) \bigg] + \frac{2 F_V^2}{ F_K  }\, D_{K^*}(t)\,\bigg[ d_3 t +
(d_1+8d_2-d_3) m_K^2 \bigg]\times\nonumber\\
& &  \left[ \frac{1}{M_\rho^2}-\frac{\mathrm{sin}^2\theta}{3M_\omega^2}
\left( 1-2\sqrt{2} \mathrm{cotg}\theta\right) -\frac{\mathrm{cos}^2\theta}{3M_\phi^2} \left( 1+2\sqrt{2} \mathrm{tg}\theta\right)\right] \,.
\end{eqnarray}

\begin{figure}[ht]
\begin{center}
\includegraphics[scale=0.65]{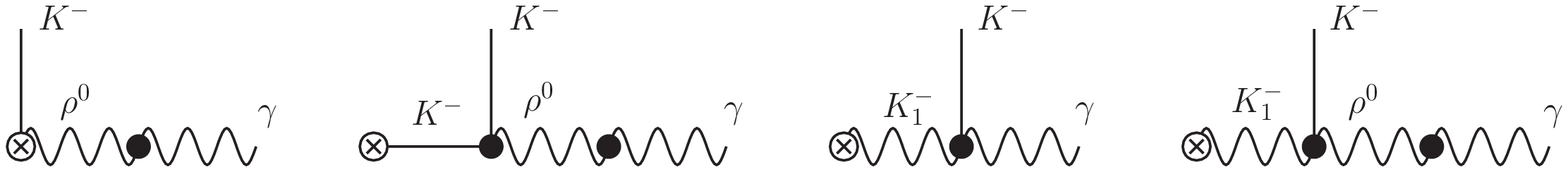}
\caption{Axial-vector current contributions to $\tau^-\rightarrow  K^- \gamma \nu_\tau$. \label{fig.k.a}}
\end{center}
\end{figure}

The axial-vector current contributions to the $SD$ part are given in Figure \ref{fig.k.a}. The corresponding analytical result is
\begin{equation}
i \mathcal{M}_{\mathrm{SD}_{A}}=  G_F\, V_{us}\, e \,\overline{u}_{\nu_\tau}(q)\,\gamma^\mu (1-\gamma_5)\,u_\tau(s)
 \,\epsilon^{\nu}(k)\, \big[ (t-m_K^2)g_{\mu\nu}-2k_\mu p_\nu \big]\, F_A^K(t)\,,
\end{equation}
where the axial-vector form-factor $F_A^K(t)$ is
\begin{eqnarray} \label{fakt}
F_A^K(t) &=& \frac{F_V^2}{4F_K }\left( 1-\frac{2G_V}{F_V}\right)
\left( \frac{1}{M_\rho^2}+\frac{\mathrm{cos}^2\theta}{M_\phi^2}+\frac{\mathrm{sin}^2\theta}{M_\omega^2} \right)
- \frac{F_A^2}{ 2 F_K} \bigg[ \mathrm{cos}^2\theta_A D_{K_{1H}}(t) +\mathrm{sin}^2\theta_A D_{K_{1L}}(t) \bigg]
\nonumber \\ & &
+ \frac{ F_A F_V}{ \sqrt2 F_K }\,\bigg[ \mathrm{cos}^2\theta_A D_{K_{1H}}(t) +\mathrm{sin}^2\theta_A D_{K_{1L}}(t) \bigg]\,
\nonumber \\ &&\qquad\quad
\times\bigg( \frac{1}{M_\rho^2}+\frac{\mathrm{cos}^2\theta}{M_\phi^2}+\frac{\mathrm{sin}^2\theta}{M_\omega^2} \bigg)
\bigg( - \lambda'' t +\lambda_0 m_K^2 \bigg)\,.
\end{eqnarray}
We have used the notations of $K_{1H}$ and $K_{1L}$ for the physical states $K_1(1400)$
and $K_1(1270)$, respectively, and the mixing angle $\theta_A$ is defined in Eq.(\ref{thetaa}) as we explain in the following.

The $K_{1A}$ state is related to the physical states $K_{1}(1270)$, $K_{1}(1400)$ through:
\begin{equation}\label{thetaa}
K_{1A} \,=\, \cos\theta_A\,\, K_{1}(1400) + \sin\theta_A \,\,K_{1}(1270) \,.
\end{equation}
The nature of $K_1(1270)$ and $K_{1}(1400)$ is not clear yet. It has been proposed in Ref.~\cite{Suzuki:1993yc} that they result from the mixing of
the states $K_{1A}$ and $K_{1B}$, where $K_{1A}$ denotes the strange partner of the axial vector resonance a$_1$ with $J^{PC}=1^{++}$ and
$K_{1B}$ is the corresponding strange partner of the axial vector resonance $b_1$ with $J^{PC}=1^{+-}$. However in this work, we will not
include the nonet of axial vector resonances with $J^{PC}=1^{+-}$ \cite{Ecker:2007us}. As argued in Ref.~\cite{Suzuki:1993yc}, the contributions
 from these resonances to tau decays are proportional to $SU(3)$ symmetry breaking effects. Moreover, as one can see later, we
will always assume $SU(3)$ symmetry for both vector and axial-vector resonances in deriving the T-matrix. For the pseudo-Goldstone bosons, physical
masses will arise through the chiral symmetry breaking mechanism in the same way as it happens in $QCD$. For the vector resonances, the
experimental values will be taken into account in the kinematics, while in the case of the axial-vector mesons, we will take the determination of the
$a_1$ mass from Ref.~\cite{Dumm:2009va} and the masses of the $K_1$ resonances from  $PDG$~\cite{PDG08}.\\

\section{Constraints from $QCD$ asymptotic behaviour} \label{QCDconstraints}

In this part, we will exploit the asymptotic results of the form factors from perturbative $QCD$ to constrain the resonance couplings. When
discussing the high-energy constraints, we will work both in chiral and $SU(3)$ limits, which indicates we will not distinguish the form
factors with pion and kaon, that are identical in this case \footnote{The results of this procedure are independent of taking the chiral limit \cite{RuizFemenia:2003hm}.}.

For the vector form factor, the asymptotic result of perturbative $QCD$ has been derived in Ref.~\cite{Lepage:1979zb, Brodsky:1981rp}
\begin{equation} \label{lbhe}
F_V^P(t \to -\infty) = \frac{F}{t}\,,
\end{equation}
where $F$ is the pion decay constant in the chiral limit. From the above asymptotic behaviour, we find three constraints on the
resonance couplings
\begin{equation} \label{asymt1}
c_1-c_2+c_5 \,=\, 0 \,,
\end{equation}
\begin{equation} \label{asymt0}
c_2-c_1+c_5-2 c_6 \,=\, \frac{\sqrt{2} N_C M_V}{32 \pi^2 F_V} + \frac{\sqrt{2} F_V}{M_V} d_3 \,,
\end{equation}
\begin{equation} \label{asymtm1}
c_2-c_1+c_5-2 c_6 \,=\, \frac{3 \sqrt{2} F^2 }{4 F_V M_V} + \frac{\sqrt{2} F_V}{M_V} d_3 \,,
\end{equation}
where the constraints in Eqs.(\ref{asymt1}), (\ref{asymt0}) and (\ref{asymtm1}) are derived from $\mathcal{O}(t^1)$, $\mathcal{O}(t^0)$
 and $\mathcal{O}(t^{-1})$, respectively. Combining the above three constraints, we have
\begin{equation}\label{asymt0new}
c_5- c_6 \,=\, \frac{N_C M_V}{32\sqrt2 \pi^2 F_V} + \frac{F_V }{\sqrt2 M_V} d_3
\end{equation}
\begin{equation} \label{asymfmv}
F=\frac{M_V \sqrt{N_C}}{2\sqrt6 \pi}\,,
\end{equation}
where the constraint of Eq.(\ref{asymfmv}) has already been noticed in~\cite{Decker:1993ut, Lepage:1979zb, Brodsky:1981rp}.

It is worthy to point out different results for the asymptotic behavior of the vector form factor $F_{\pi\gamma}(t)$ have
also been noted in different frameworks, such as the ones given in Refs.~\cite{Manohar:1990, Gerard:1995}. In Refs.~\cite{Lepage:1979zb, Brodsky:1981rp},
the result was obtained in the parton picture and unavoidably the parton distribution
function for the pion has to be imposed to give the final predictions.  $OPE$ technique was exploited to
obtain its prediction in \cite{Manohar:1990}, which led to the conclusion that potentially large $QCD$ corrections could exist.
In \cite{Gerard:1995}, the form factor was discussed by using Bjorken-Johnson-Low theorem~\cite{BJL}. Variant methods have also been
used to analyze this form factor: the corrections from the transverse momentum of the parton were addressed in Ref.~\cite{Maboqiang:1996}; a
$QCD$ sum rule method was applied to derive the asymptotic behavior in Ref.~\cite{Radyushkin:1996}.
In the present discussion, we focus our attention on Refs.~\cite{Manohar:1990, Gerard:1995}, while
the study for the other results can be done analogously.
Although different results agree with the same leading power of the square momentum for large $t$, behaving as $1/t$,
they predict different coefficients, such as
\bqa \label{otherhe1}
F_V^P(t \to -\infty)  &= & \frac{2F}{3t}\,,
\eqa
from Ref.~\cite{Manohar:1990} and
\bqa\label{otherhe2}
 F_V^P(t \to -\infty)  &=&   \frac{F}{3t}\,.
\eqa
from Ref.~\cite{Gerard:1995}.

By doing the same analyses as we have done by using the Lepage-Brodsky result in Eq.(\ref{lbhe}) to constrain
the resonance couplings, we can straightforwardly
get the constraints from the short-distance behaviors given in Eqs.(\ref{otherhe1})-(\ref{otherhe2}). Apparently the matching results given
in Eqs.(\ref{asymt1})-(\ref{asymt0}) will stay the same, since they are derived from $\mathcal{O}(t^1)$ and $\mathcal{O}(t^0)$.
Comparing with the result of Eq.(\ref{asymtm1}) from the matching of $\mathcal{O}(t^{-1})$ by using the coefficient in Eq.(\ref{lbhe}),
the corresponding results by using Eqs.(\ref{otherhe1}) and (\ref{otherhe2}) are respectively
\bqa \label{asymtm1other}
c_2-c_1+c_5-2 c_6 \, &=&\, \frac{ F^2 }{ 2 \sqrt{2} F_V M_V} + \frac{\sqrt{2} F_V}{M_V} d_3 \,,   \\
c_2-c_1+c_5-2 c_6 \, &=&\, \frac{  F^2 }{4\sqrt{2} F_V M_V} + \frac{\sqrt{2} F_V}{M_V} d_3 \,,
\eqa
which lead to the following results, in order, by combining Eqs.(\ref{asymt1})-(\ref{asymt0})
\bqa \label{asymfmvother1}
F=\frac{M_V \sqrt{N_C}}{4\pi}\,,    \\
F=\frac{M_V \sqrt{N_C}}{2\sqrt2 \pi}\,.  \label{asymfmvother2}
\eqa

The formulae displayed in Eq.(\ref{asymfmv}), Eq.(\ref{asymfmvother1}) and  Eq.(\ref{asymfmvother2}) provide a simple way to
discriminate between different asymptotic behaviors.
The chiral limit values for the pion decay constant $F$ and the mass of the lowest vector multiplet $M_V$ have been
thoughtfully studied at the leading order of $1/N_C$ in Ref.~\cite{guo:2009}, which predicts $F= 90.8$ MeV and $M_V= 764.3$ MeV.
The different results shown in Eq.(\ref{asymfmv}), Eq.(\ref{asymfmvother1}) and  Eq.(\ref{asymfmvother2})
 from  matching  different short-distance behaviors of Refs.~\cite{Lepage:1979zb,Manohar:1990,Gerard:1995}
deviate from the phenomenology study at the level of $5\%, 16\%, 64\%$ respectively \footnote{The results are mildly changed when estimating the values
of $F$ and $M_V$ by  $F_\pi$ and $M_\rho$ respectively, as expected.}, which implies that the short-distance behavior in
Eq.(\ref{lbhe}) is more reasonable than the ones in Eqs.(\ref{otherhe1})-(\ref{otherhe2}).
Hence for the matching result at the order of $1/t$, we will use the one in Eq.(\ref{asymfmv}) throughout the following discussion.
However we stress the inclusion of extra multiplets of vector resonances or the sub-leading corrections in $1/N_C$ may alter the
current conclusion.

The high-energy constraints on the resonance couplings $c_i$ and $d_i$
have been studied in different processes. The $OPE$ analysis of the $VVP$ Green Function gives~\cite{RuizFemenia:2003hm}
\begin{equation}\label{asymvvpc5}
c_5- c_6 \,=\, \frac{N_C M_V}{64\sqrt2 \pi^2 F_V} \,,
\end{equation}
\begin{equation}\label{asymvvpd3}
d_3 \,=\, -\frac{N_C M_V^2}{64 \pi^2 F_V^2}+\frac{F^2}{8F_V^2}\,.
\end{equation}

The constraint from $\tau^- \to (V P)^- \nu_\tau$ study leads to
\begin{equation}\label{asymtauvp}
c_5- c_6 \,=\, -\frac{F_V }{\sqrt2 M_V} d_3\,,
\end{equation}
if one neglects the heavier vector resonance multiplet~\cite{Guo:2008sh}.

The results from the analysis of $\tau^- \to (K K \pi)^- \nu_\tau$ are~\cite{Roig:2007yp}
\begin{eqnarray} \label{asymtaukkpi}
c_5- c_6 \,=\, \frac{N_C M_V F_V}{192\sqrt2 \pi^2 F^2} \,, \nonumber \\
d_3 \,=\, -\frac{N_C M_V^2}{192 \pi^2 F^2}\,.
\end{eqnarray}
It is easy to check that the results of Eqs.(\ref{asymtauvp}) and (\ref{asymtaukkpi}) are consistent. Combining Eqs.(\ref{asymt0new})
and (\ref{asymtauvp}) leads to
\begin{eqnarray} \label{asym_taupg_tauvp}
c_5- c_6 &=& \frac{N_C M_V}{64\sqrt2 \pi^2 F_V} \,,\nonumber \\
d_3 &=& -\frac{N_C M_V^2}{64 \pi^2 F_V^2}\,,
\end{eqnarray}
where the constraint of $c_5 -c_6$ is consistent with the result from the $OPE$ analysis of the $VVP$ Green Function~\cite{RuizFemenia:2003hm},
 while the result of $d_3$ is not.

By demanding the consistency of the constraints derived from the processes of $\tau^- \to P^- \gamma \nu_\tau$ and
$\tau^- \to (V P)^- \nu_\tau$ given in Eq.(\ref{asym_taupg_tauvp}) and the results from $\tau^- \to (K K \pi)^- \nu_\tau$ given in
Eq.(\ref{asymtaukkpi}), we get the following constraint
\begin{equation}\label{ksrf3fv}
F_V=\sqrt3 F\,.
\end{equation}
If one combines the high-energy constraint from the two pion vector form factor~\cite{Ecker:1989yg}
\begin{equation}
F_V G_V = F^2\,,
\end{equation}
and the result of Eq.(\ref{ksrf3fv}) we get here, the modified Kawarabayashi-Suzuki-Riazuddin-Fayyazuddin ($KSRF$) relation \cite{Kawarabayashi:1966kd, Riazuddin:1966sw}
is derived
\begin{equation}
F=\sqrt3 G_V\,,
\end{equation}
which is also obtained in the partial wave dispersion relation analysis of $\pi \pi$ scattering by properly including the contributions from the crossed channels
 ~\cite{Guo:2007ff}\,.

Although the branching ratios for the modes of $\tau \to P \gamma \nu_\tau$ which we are discussing could be higher than for some modes that have been already detected,
they have
not been observed yet. Lacking of experimental data, we will make some theoretically and phenomenologically based assumptions in order to
present our predictions for the spectra and branching ratios.

Taking into account the previous relations one would have $F_V^\pi(t)$ in terms of $c_1+c_2+8c_3-c_5$ and $d_1+8d_2-d_3$.
For the first combination, $c_1+c_2+8c_3-c_5=c_1+4c_3$ ($c_1-c_2+c_5 = 0$ has been used), the prediction for $c_1+4c_3$ in \cite{RuizFemenia:2003hm} yields
$c_1+c_2+8c_3-c_5=0$. In Ref.~\cite{RuizFemenia:2003hm} the other relevant combination of couplings is also restricted: $d_1+8d_2-d_3=\frac{F^2}{8F_V^2}$. In
$F_V^K(t)$, $c_4$ appears in addition. There is a phenomenological determination of this coupling in the study of the $KK\pi$ decay modes
of the $\tau$~\cite{Dumm:2009kj}: $c_4=-0$.$07\pm0$.$01$.

Turning now to the axial-vector form factor, it still depends on four couplings in both channels: $F_A$, $M_A$, $\lambda''$
and $\lambda_0$. If one invokes the once subtracted dispersion relation for the axial vector form factor, as done in Ref.~\cite{Decker:1993ut}, one
can not get any constraints on the resonance couplings from the axial vector form factors given in Eqs.(\ref{fapit}) and (\ref{fakt}). In
fact by demanding the form factor to satisfy the unsubtracted dispersion relation, which guarantees a better high-energy limit, we can get
the following constraint
\begin{equation}
 \lambda''=\frac{2G_V-F_V}{2\sqrt{2}F_A}\,,
\end{equation}
which has been already noted in ~\cite{Cirigliano:2004ue}.

In order to constrain the free parameters as much as possible, we decide to exploit the constraints from the Weinberg sum
rules ($WSR$) \cite{Weinberg:1967kj}: $F_V^2-F_A^2=F^2$ and $M_V^2F_V^2-M_A^2F_A^2=0$, yielding
\begin{equation}
 F_A=2F^2\;\;,\;M_A=\frac{6\pi F}{\sqrt{N_C}}\,.
\end{equation}
For the axial vector resonance coupling $\lambda_0$, we use the result from Ref.~\cite{Dumm:2009va, Cirigliano:2004ue}
\begin{equation}
\lambda_0=\frac{G_V}{4\sqrt{2}F_A }\,.
\end{equation}
To conclude this section, we summarize the previous discussion on the high-energy constraints
\begin{eqnarray}\label{heconstraint}
&&F_V=\sqrt3 F \,,\quad G_V=\frac{F}{\sqrt3}\,,\quad F_A=\sqrt2 F\,,\quad M_V=\frac{2\sqrt6 \pi F}{\sqrt{N_C}} \,,\quad
M_A=\frac{6\pi F}{\sqrt{N_C}}\,,
\nonumber \\ &&
\lambda_0=\frac{1}{8\sqrt3}\,,\quad \lambda''=-\frac{1}{4\sqrt3}\,,\quad
c_5-c_6=\frac{\sqrt{N_C}}{32\pi}\,,\quad d_3=-\frac{1}{8}\,.
\end{eqnarray}
In the above results, we have discarded the constraint in Eq.(\ref{asymvvpd3}), which is the only inconsistent result with the others.

\section{Phenomenological discussion} \label{Pheno}

Apart from the parameters we mentioned in the last section, there is still one free coupling $\theta_A$, which
describes the mixing of the strange axial vector resonances in Eq.(\ref{thetaa}). The value of $\theta_A$ has already been
determined in the literature~\cite{Guo:2008sh, Suzuki:1993yc, Cheng:2003bn}. We recapitulate the main results in the following.

In Ref.~\cite{Suzuki:1993yc}, it has given $\theta_A\sim33^\circ$. In Ref.~\cite{Guo:2008sh}, $|\theta_A| \sim 58.1^\circ$
is determined through the considered decays $\tau^-\to (VP)^-\nu_\tau$. In Ref.~\cite{Cheng:2003bn},
the study of $\tau\to K_1 \nu_\tau$ gives $|\theta_A|=^{37^\circ}_{58^\circ}$ as the two possible solutions. The decay $D\to K_1 \pi$ allows to
conclude that $\theta_A$ must be negative and it is pointed out that the observation of $D^0\to K_1^-\pi^+$ with a branching ratio $\sim5\cdot10^{-4}$
would imply $\theta_A\sim-58^\circ$. However, a later analysis in Ref.~\cite{Cheng:2007mx} finds that the current measurement of $\bar{B}^0\to K_1^-(1400) \pi^+$
\cite{PDG08} favors a mixing angle of $-37^\circ$ over $-58^\circ$. In this respect, the relation
\begin{equation}
 \Big|\Gamma\left(J/\Psi\to K_1^0(1400)\overline{K}^0\right)\Big|^2\,=\,\mathrm{tg}\theta_A^2\,\Big|\Gamma\left(J/\Psi\to K_1^0(1270)\overline{K}^0\right)\Big|^2
\end{equation}
would be very useful to get $\theta_A$, once these modes are detected. In the following discussion, we will show the results using
both $|\theta_A|=37^\circ$ and $58^\circ$. The other inputs are given in the Appendix.\ref{appendix-input}.

\subsection{Results only with $WZW$ contribution}
As it was stated before it is strange that the decay modes $\tau \to P \gamma \nu_\tau$ have not been detected
 so far. The most naive and completely model independent estimate would be to just include the $IB$ part and the $WZW$ contribution to the $VV$ part, as the latter
is completely fixed by $QCD$. We know that doing this way we are losing the contribution of vector and axial-vector resonances, that should be important in
the high-$x$ region. However, even doing so one is able to find that the radiative decay $\t^-\to\pi^-\gamma \nu_\t$ has a decay probability larger than
the mode $\t^-\to K^+K^-K^-\nu_\t$ \footnote{$\Gamma\left(\t^-\to K^+K^-K^-\nu_\t\right)\,=\,3$.$579(66)\cdot 10^{-17}$ GeV.}.
For a reasonably low cut-off on the photon energy this conclusion holds for the $\t^-\to K^-\gamma \nu_\t$ as well.

\begin{figure}[ht]
\begin{center}
\includegraphics[angle=0, width=1.0\textwidth]{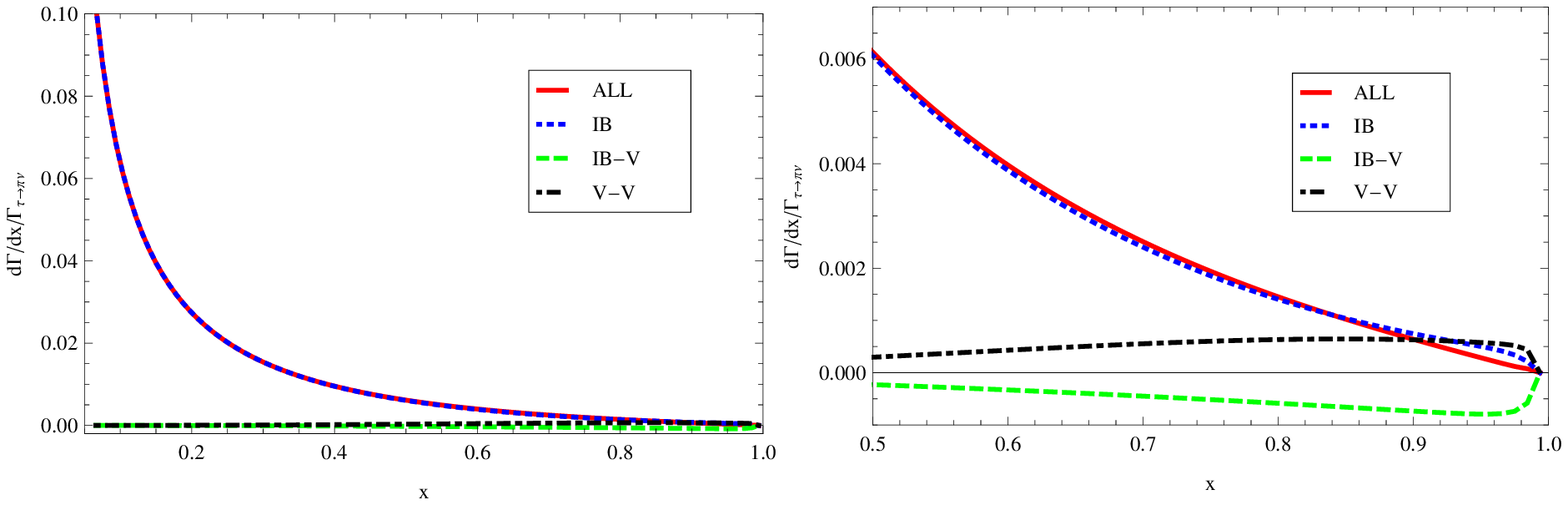}
\caption{ Differential decay width (in units of $\Gamma_{\tau \to \pi \nu_\tau}$ ) of the process $\tau^-\rightarrow  \pi^- \gamma \nu_\tau$
including only the model independent contributions  as a function of $x$ . For the form factors only the $WZW$ term is considered for this estimate,
where the axial vector contribution is absent.  The right plot is the close-up of the left one in region of $x>0.5$.  \label{fig.piwzwX}}
\end{center}
\end{figure}

\begin{figure}[ht]
\begin{center}
\includegraphics[angle=0, width=1.0\textwidth]{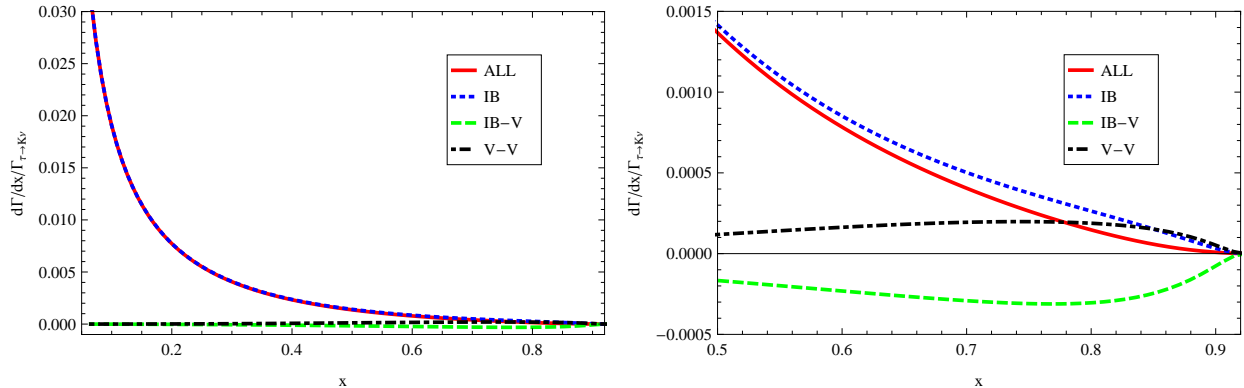}
\caption{ Differential decay width (in units of $\Gamma_{\tau \to K \nu_\tau}$ ) of the process $\tau^-\rightarrow  K^- \gamma \nu_\tau$ including only the model
independent contributions as a function of $x$ . For the form factors only the $WZW$ term is considered for this estimate, where the axial vector contribution is absent.
 The right plot is the close-up of the left one in region of $x>0.5$.  \label{fig.kwzwX}}
\end{center}
\end{figure}

Before seeing this, we will discuss briefly the meaning of cutting on the photon energy \footnote{A cut on the photon energy was introduced in Sect.
\ref{Kinematics}.}. As it is well known \cite{Kinoshita:1962ur, Peskin:1995ev} the $IR$ divergences due to the vanishing photon mass cancel
when considering at the same time the non-radiative and the radiative decays with one photon \footnote{In general, the $IR$ divergences of the n-photon decay
are canceled by those in the $n+1$-photon process.}. In practice, this translates into mathematical language the physical notion that
the detectors have a limited angular resolution that defines a threshold detection angle for photons. If one considers a photon emitted with a smaller angle it
should be counted together with the non-radiative decay as it is effectively measured in this way. The sum is of course an $IR$ safe observable. The splitting depends
on the particular characteristics of the experimental setting. Obviously, the branching fraction for the radiative decay depends on this cut-off energy. We will consider
here the case $E_{\gamma\,\mathrm{thr}}\,=\,50$ MeV, that corresponds to $x \, = \, 0$.$0565$. In order to illustrate the dependence on this variable, we will also show
 the extremely conservative case of  $E_{\gamma\,\mathrm{thr}}\,=\,400$ MeV ($x \, = \, 0$.$45$). In the first case we obtain $\Gamma\left(
\t^-\to\pi^-\gamma \nu_\t\right)\,=\,3.182\times 10^{-15}$ GeV, and in the second one we are still above the $3K$ decay width, $\Gamma\left(
\t^-\to\pi^-\gamma \nu_\t\right)\,=\,3.615\times  10^{-16}$ GeV. In Figure~\ref{fig.piwzwX} we plot the photon spectrum of $\tau^- \to \pi^- \gamma \nu_\tau$.
Proceeding analogously for the decay with a $K^-$, we find: $\Gamma\left(\t^-\to K^-\gamma
\nu_\t\right)\,=\,6.002 \times 10^{-17}$ GeV for $E_{\gamma\,\mathrm{thr}}\,=\,50$ MeV, and $\Gamma\left(\t^-\to K^-\gamma \nu_\tau \right)
\,=\,4.589 \times 10^{-18}$ GeV for $E_{\gamma\,\mathrm{thr}}\,=\,400$ MeV. The photon spectrum of  $\tau^- \to K^- \gamma \nu_\tau$ can be seen in Figure~\ref{fig.kwzwX}.
 For any reasonable cut on $E_{\gamma}$ these modes should have already been detected by the $B$-factories.

Already at this level of the phenomenological analysis, the question of the accuracy on the detection of soft photons at $B$-factories \cite{Becirevic:2009aq} arises
\footnote{See however, Ref.\cite{Guo:2010ny}}.
An error larger than expected (here and in some undetected particle interpreted as missing energy, in addition to a gaussian treatment
 of systematic errors) could enlarge the uncertainty claimed on the measurement of $B^-\to\tau^- \nu_\tau$ \cite{PDG08} when combining the
$Belle$~\cite{Ikado:2006un} and $BaBar$ measurements~\cite{Aubert:2007bx, Aubert:2007xj} taking it closer to the Standard Model expectations.

\subsection{ Results with the resonance contributions for the pion channel}

Next we include also the model-dependent contributions. Since in the kaon channel there are uncertainties associated with the off-shell width of
the strange axial-vector resonance and the mixing of the corresponding light and heavy states, we will present first the pion channel where there are not any uncertainties of these types and everything is fixed.

\begin{figure}[ht]
\begin{center}
\includegraphics[angle=0, width=1.0\textwidth]{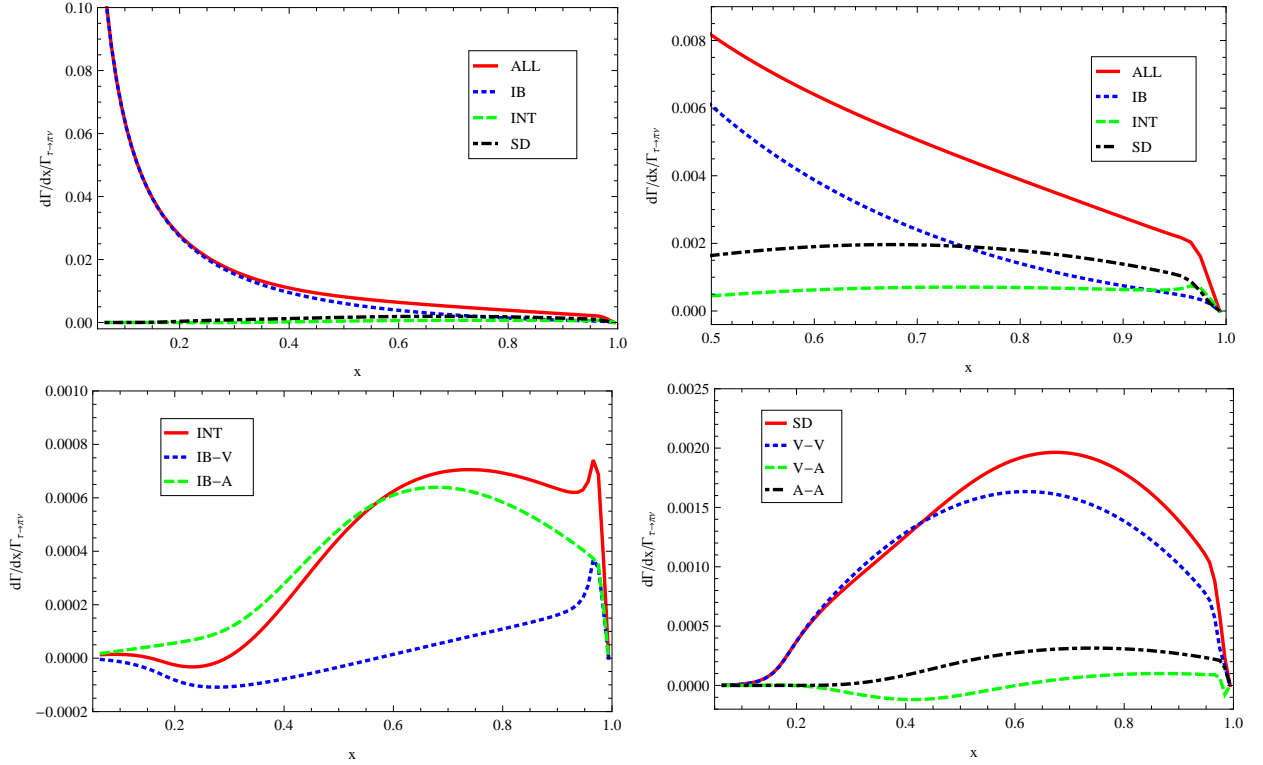}
\caption{ Differential decay width ( in units of $\Gamma_{\tau \to \pi \nu_\tau}$ ) of the process $\tau^-\rightarrow \pi^- \gamma \nu_\tau$
including all contributions as a function of $x$ . The top right plot is the close-up of the top left one in region of $x>0.5$.
The bottom left and right plots are to display the compositions of the interference and structure dependent parts, respectively.  \label{fig.pitot4X}}
\end{center}
\end{figure}

In Fig.\ref{fig.pitot4X} the resulting photon spectrum of the process $\tau^-\rightarrow  \pi^- \gamma \nu_\tau$ is
displayed. In order to display clearly how the different parts contribute to the spectrum, we have given the close-up of the spectrum for the high $x$ region and also shown
the separate contributions in Fig.\ref{fig.pitot4X}. For ''soft'' photons ($x_0\lesssim0.3$) the
internal bremsstrahlung dominates completely. One should note that for very soft photons the multi-photon production rate becomes important, thus making that
our $\cO(\alpha)$ results are not reliable too close to the $IR$ divergence $x\,=\,0$. We qualitatively agree with the results in $DF$ papers,
for the same order of $\alpha$ to the three significant figures shown in Ref. \cite{Decker:1993ut}.

The spectrum is significantly enhanced by $SD$ contributions for hard photons ($x_0\gtrsim0.4$), as we can see in Fig.\ref{fig.pitot4X}. From Fig.\ref{fig.pitot4X}
one can also see that the vector current contribution mediated by the vector resonances dominates the $SD$ part. The interference terms between the bremsstrahlung
and the  $SD$ parts are also shown in Fig.\ref{fig.pitot4X}. If we compare the predicted curves with those in Ref. \cite{Decker:1993ut}
we see that the qualitative behaviour is similar: the $IB$ contribution dominates up to $x\sim0.75$. For larger photon energies, the $SD$ part is predominantly
due to the $VV$ contribution and overcomes the $SI$ part. We confirm the peak and shoulder structure shown at $x\sim1$ in the interference contribution, which is
essentially due to $IB-V$ term, and also the $V-A$ term, that is in any case tiny.

\begin{figure}[ht]
\begin{center}
\includegraphics[angle=0, width=1.0\textwidth]{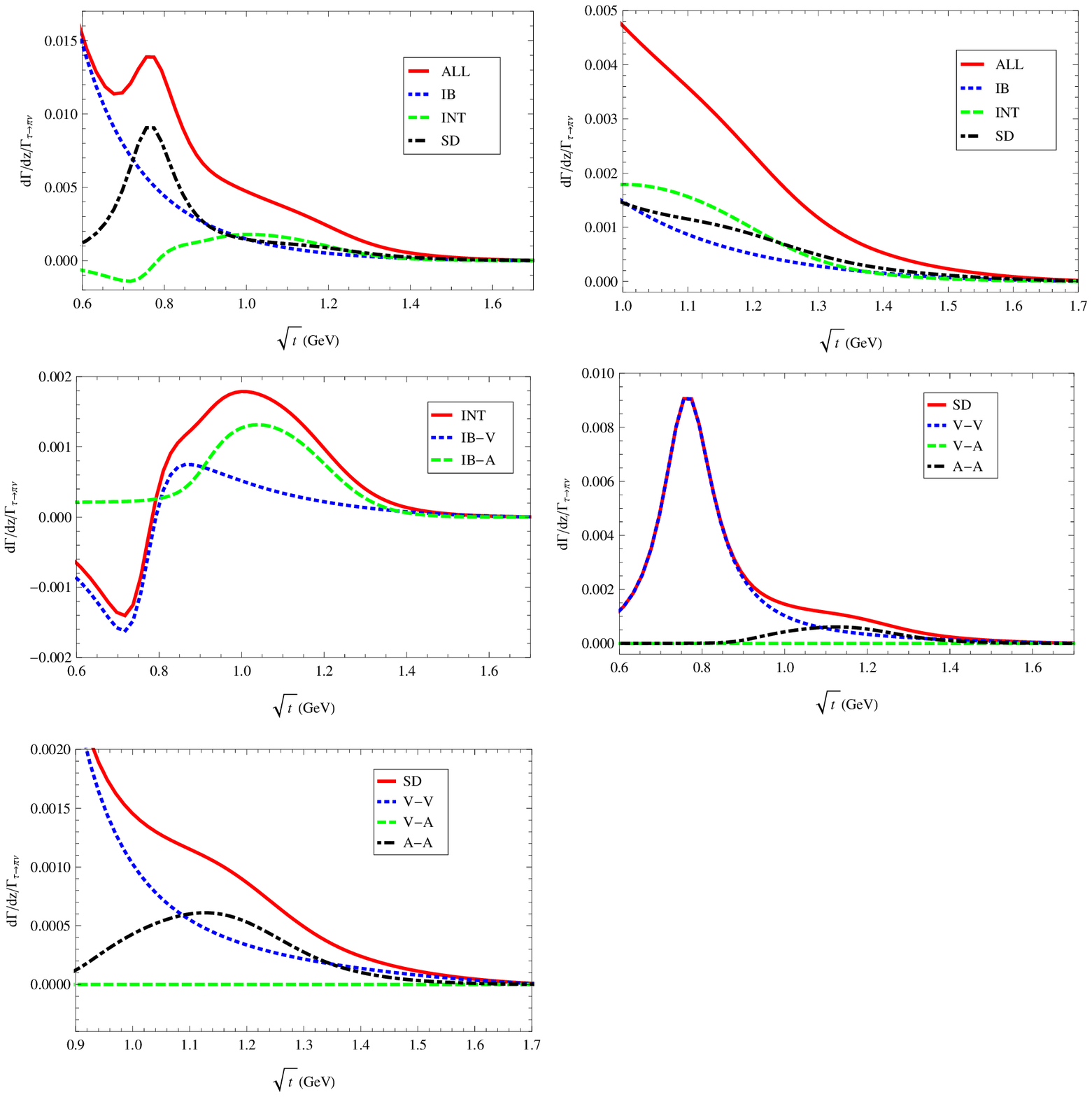}
\caption{ Differential decay width ( in units of $\Gamma_{\tau \to \pi \nu_\tau}$ ) of the process $\tau^-\rightarrow \pi^- \gamma \nu_\tau$
including all contributions as a function of $\sqrt t$. The top right plot is the close-up of the top left one in region of $\sqrt t>1.0$.
The middle left and right plots are to display the compositions of the interference and structure dependent parts, respectively.
The bottom plot is the close up of the middle right one in the region of $\sqrt t > 0.9$ GeV.  \label{fig.pitot4Z}}
\end{center}
\end{figure}

While the integration over the $IB$ part needs an $IR$ cut-off, the $SD$ part does not. We have performed the integration over the complete phase
space, yielding \footnote{Here and in what follows, all contributions to the partial decay width are given in units of the non-radiative decay.}:
\begin{equation} \label{Gamma_SD_pi}
 \Gamma_{VV}=0.99\cdot10^{-3}\;,\quad\Gamma_{VA}\sim0\;,\quad\Gamma_{AA}=0.15\cdot10^{-3}\,\Rightarrow \Gamma_{SD}=1.14\cdot10^{-3}\,.
\end{equation}
 Our number for $\Gamma_{SD}$ lies between the results for the monopole and tripole parametrizations in Ref. \cite{Decker:1993ut}. However, they get a smaller(larger)
$VV$($AA$) contribution than we do by $\sim20\%$($\sim200\%$). This last discrepancy is due to the off-shell a$_1$ width they use. In fact, if we use the constant width
approximation we get a number very close to theirs for the $AA$ contribution. With the understanding of the a$_1$ width in the
$\tau\to3\pi\nu_\tau$ observables \cite{Dumm:2009va}, we
can say that their (relatively) high $AA$ contribution is an artifact of the ad-hoc off-shell width used. Since the numerical difference in varied vector off-shell widths
is not that high, the numbers for $VV$ are closer.

The numbers in Eq.(\ref{Gamma_SD_pi}) are translated into the following branching ratios
\begin{equation} \label{BR_SD_pi}
 \mathrm{BR}_{VV}\left(\tau\to\pi\gamma\nu_\tau\right)=1.05\cdot10^{-4}\;,\quad\mathrm{BR}_{AA}\left(\tau\to\pi\gamma\nu_\tau\right)=0.15\cdot10^{-4}\,.
\end{equation}
We can also compare the $VV$ value with the narrow-width estimate. Taking into account the lowest-lying resonance $\rho$(770) we get
\begin{eqnarray} \label{est_VV_pi}
 \mathrm{BR}_{VV}\left(\tau\to\pi\gamma\nu_\tau\right)& \sim & BR(\tau\to\rho\nu_\tau)\times BR(\rho\to\pi\gamma)\sim
 BR(\tau\to\pi^-\pi^0\nu_\tau)BR(\rho\to\pi\gamma)\nonumber\\
& \sim & 25.52\%\times4.5\cdot10^{-4}=1.15\cdot10^{-4}\,,
\end{eqnarray}
which is quite a good approximation.

In Table \ref{Tabledifferentcontributionspi} we display (for two different values of the photon energy cut-off) how the different parts contribute to
the total rate. For a low-energy cut-off the most of the rate comes from $IB$ while for a higher-energy one the $SD$ parts (particularly the $VV$ contribution)
gain importance. While the $VA$ contribution is always negligible, the $IB-V$, $IB-A$ and the remaining $SD$ parts ($VV$ and $AA$) have some relevance
for a higher-energy cut-off.
\begin{table}[h!]
\begin{center}
\renewcommand{\arraystretch}{1.2}
\begin{tabular}{|c|c|c|}
\hline
&$x_0=0.0565$&$x_0=0.45$\\
\hline
$IB$&$13.09\cdot10^{-3}$&$1.48\cdot10^{-3}$\\
$IB-V$&$0.02\cdot10^{-3}$&$0.04\cdot10^{-3}$\\
$IB-A$&$0.34\cdot10^{-3}$&$0.29\cdot10^{-3}$\\
$VV$&$0.99\cdot10^{-3}$&$0.73\cdot10^{-3}$\\
$VA$&$ \sim 0 $&$0.02\cdot10^{-3}$\\
$AA$&$0.15\cdot10^{-3}$&$0.14\cdot10^{-3}$\\
\hline
$ALL$&$14.59\cdot10^{-3}$&$2.70\cdot10^{-3}$\\
\hline
\end{tabular}
\caption{\small{Contributions of the different parts to the total rate in the decay $\tau^-\to\pi^-\gamma\nu_\tau$ (in units of $\Gamma_{\tau \to \pi \nu}$),
using two different cut-offs for the photon energy: $E_\gamma=50$ MeV ($x_0=0.0565$) and $E_\gamma=400$ MeV ($x_0=0.45$).}}
\label{Tabledifferentcontributionspi}
\end{center}
\end{table}

In Fig.\ref{fig.pitot4Z} we show the pion-photon invariant spectrum. We find a much better signal of the $SD$ contributions as compared with the photon
spectrum in the previous Fig.\ref{fig.pitot4X}, which has already been noticed in Ref.\cite{Decker:1993ut}. Then, the pion-photon spectrum
is better suited to study the $SD$ effects. In this case, the $VA$ part is identically zero, since this interference vanishes in the invariant mass spectrum
after integration over the other kinematic variable. Of course, in the $VV$ spectrum we see the shape of the $\rho$ contribution neatly, as it is shown in
Fig.\ref{fig.pitot4Z}, where on the contrary the a$_1$ exchange in $AA$ has a softer and broader effect. The $IB-SD$ radiation near the $a_1$ is
dominated by $IB-A$, which gives the positive contribution to the decay rate. While near the energy region of the $\rho$ resonance, we find the $IB-SD$
contribution to be negative as driven by $IB-V$ there. In the whole spectrum only the $\rho$ resonance manifests as a peak and one can barely see the signal
of $a_1$, mainly due to its broad width.

\subsection{ Results with the resonance contributions for the kaon channel}
Next we turn to the $\tau^-\rightarrow  K^- \gamma \nu_\tau$ channel. In this case, there are several sources of uncertainty that make our
prediction less controlled than in the $\tau^-\rightarrow  \pi^- \gamma \nu_\tau$ case. We comment them in turn.

Concerning the vector form factor contribution, there is no uncertainty associated with the off-shell widths of the vector resonances ,
that are implemented as we explained before. However we have observed that
the $VV$ contribution to the decay rate is much larger (up to one order of magnitude,
even for a low-energy cut-off) than the $IB$ one for $c_4\sim-0.07$,
a feature that is unexpected. In this case, one would also see a prominent bump in the spectrum, contrary to the typical monotonous fall driven by the $IB$ term.
For smaller values of $|c_4|$ this bump reduces its magnitude and finally disappears. One should also not forget that the inclusion of a second multiplet of
 resonances may vary this conclusion.

The uncertainty in the axial-vector form factors is twofold: on one side there is a broad band of allowed values for $\theta_A$, as discussed
at the beginning of this section. On the other hand, since we have not performed the analyses of the decay $\tau\to K\pi\pi\nu_\tau$ modes yet, we do not have an
off-shell width derived from a Lagrangian for the $K_{1A}$ resonances. In the $\tau\to3\pi\nu_\tau$ decays, $\Gamma_{\mathrm{a}_1}$ has the starring role~\cite{Dumm:2009va}. Since the
$K_{1A}$ meson widths are much smaller ($90\pm20$ MeV and $174\pm13$ MeV, for the $K_{1}(1270)$ and $K_{1}(1400)$, respectively) and they are hardly close to
the on-shell condition, a rigorous description of the width is not an unavoidable ingredient for a reasonable estimate. We decided to
use the expression inspired by the $\rho$ width from Ref.\cite{GomezDumm:2000fz}. The explicit forms of the $K_{1A}$ off-shell widths which we follow are given
in Appendix.\ref{appendix-width}.

\begin{figure}[ht]
\begin{center}
\includegraphics[angle=0, width=1.0\textwidth]{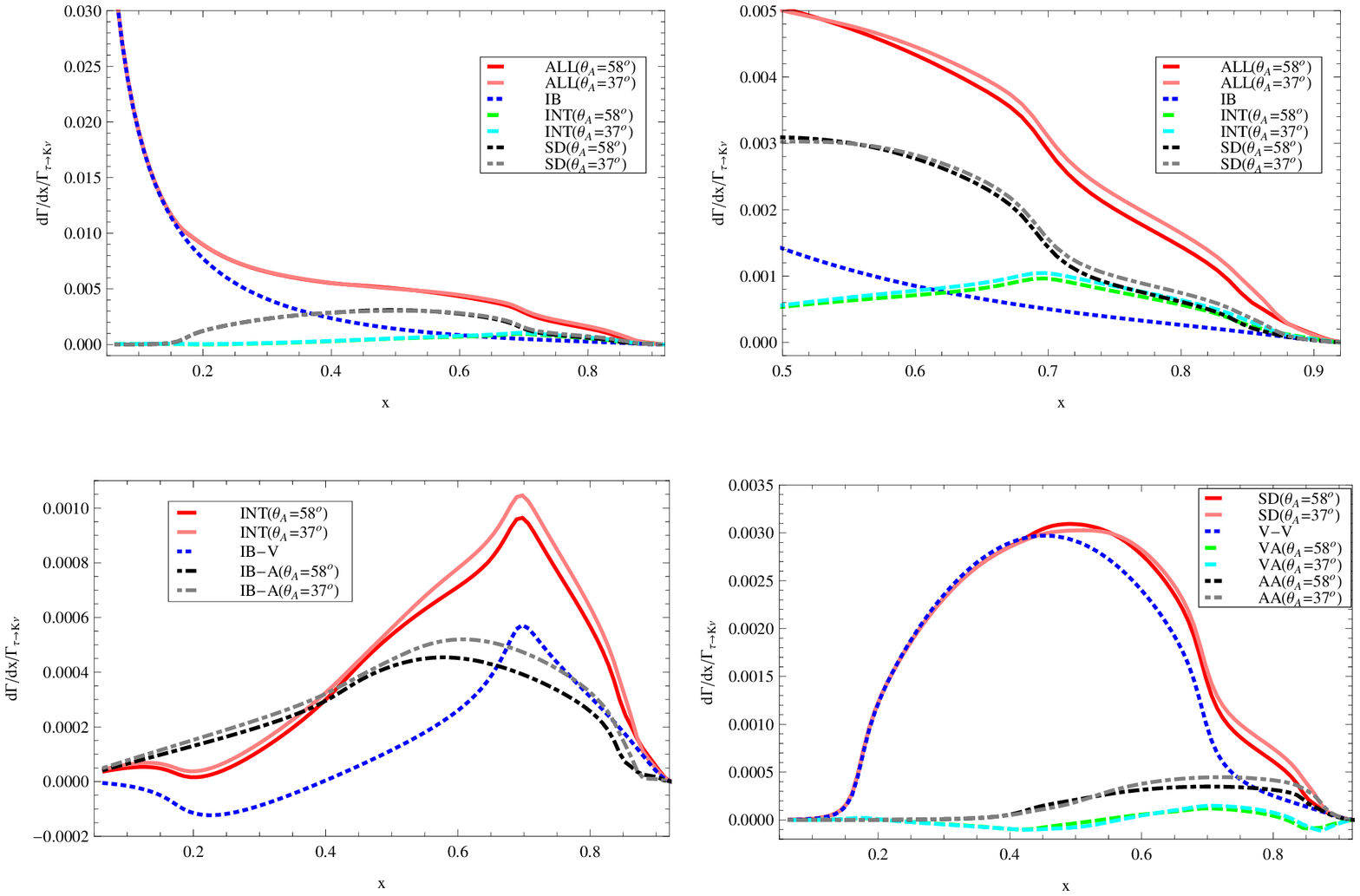}
\caption{ Differential decay width ( in units of $\Gamma_{\tau \to K \nu_\tau}$ ) of the process $\tau^-\rightarrow K^- \gamma \nu_\tau$
including all contributions as a function of $x$ with $c_4=0$. The top right plot is the close-up of the top left one in region of $x>0.5$.
The bottom left and right plots are to display the compositions of the interference and structure dependent parts, respectively.  \label{fig.ktot4Xc40}}
\end{center}
\end{figure}

\begin{figure}[ht]
\begin{center}
\includegraphics[angle=0, width=1.0\textwidth]{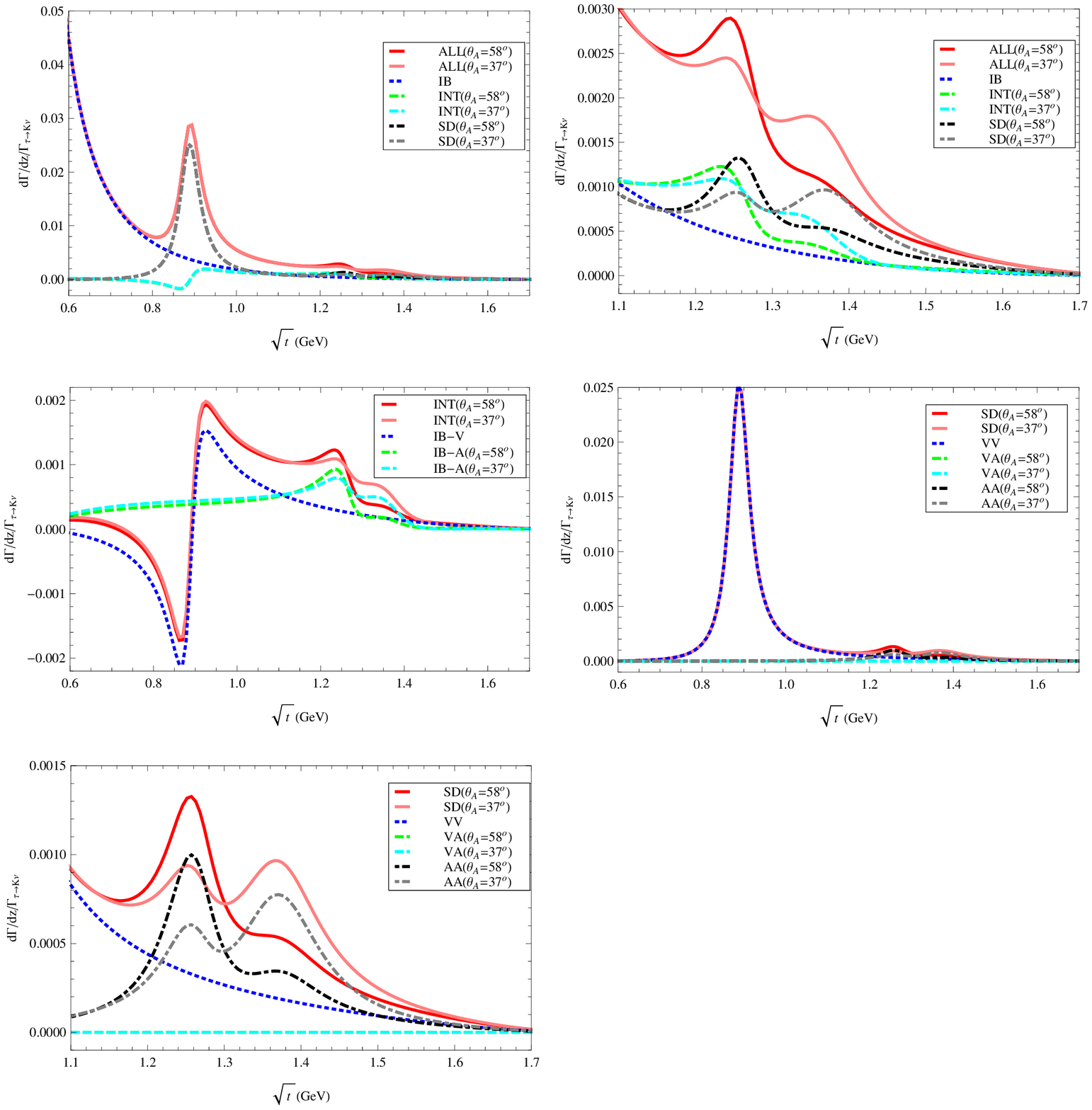}
\caption{ Differential decay width ( in units of $\Gamma_{\tau \to K \nu_\tau}$ ) of the process $\tau^-\rightarrow K^- \gamma \nu_\tau$
including all contributions as a function of $\sqrt t$ with $c_4=0$. The top right plot is the close-up of the top left one in region of $\sqrt t>1.1$ GeV.
The middle left and right plots are to display the compositions of the interference and structure dependent parts, respectively.
The bottom plot is the close up of the middle right one in the region of $\sqrt t > 1.1$ GeV.  \label{fig.ktot4Zc40}}
\end{center}
\end{figure}

\begin{figure}[h]
\begin{center}
\includegraphics[angle=0, width=1.0\textwidth]{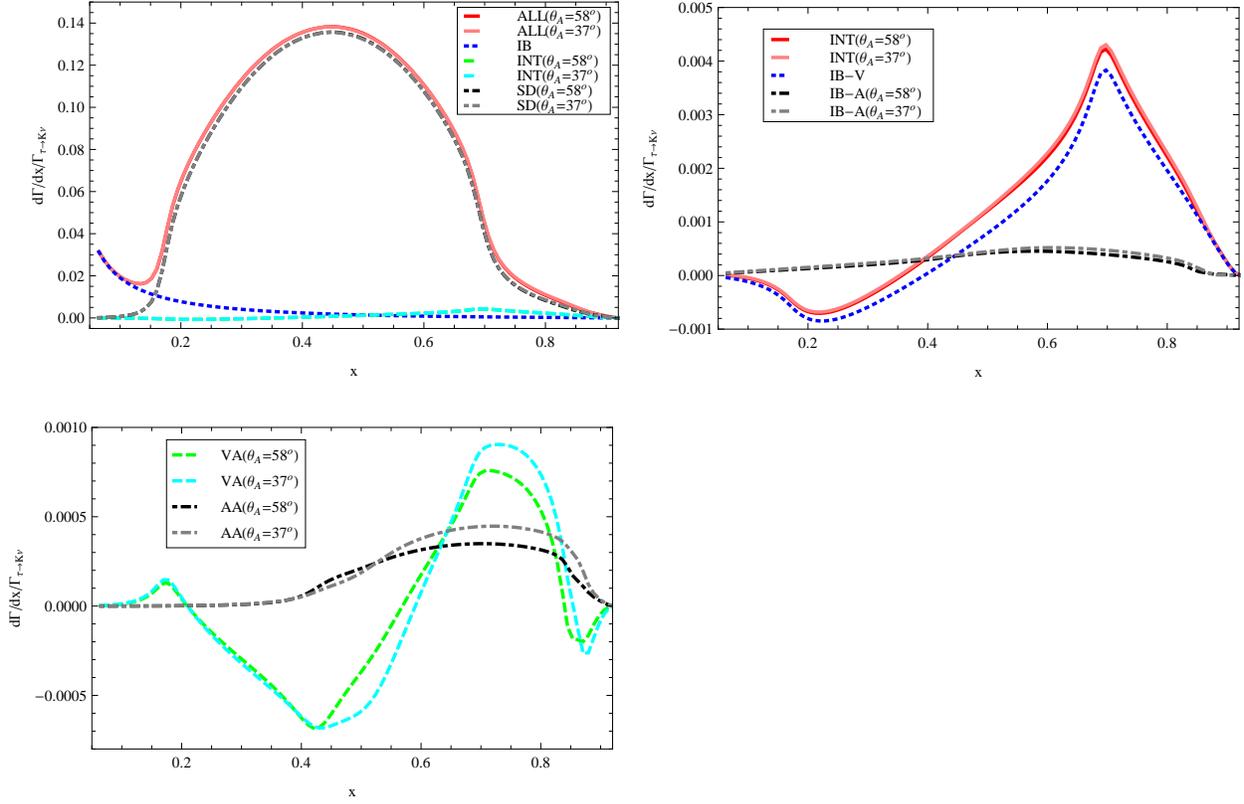}
\caption{ Differential decay width ( in units of $\Gamma_{\tau \to K \nu_\tau}$ ) of the process $\tau^-\rightarrow K^- \gamma \nu_\tau$
including all contributions as a function of $x$ with $c_4=-0.07$. The top right plot is to show the compositions
of the interference parts. The bottom left one is to show the $VA$ and $AA$ contributions. \label{fig.ktot4Xc4007}}
\end{center}
\end{figure}

\begin{figure}[h]
\begin{center}
\includegraphics[angle=0, width=1.0\textwidth]{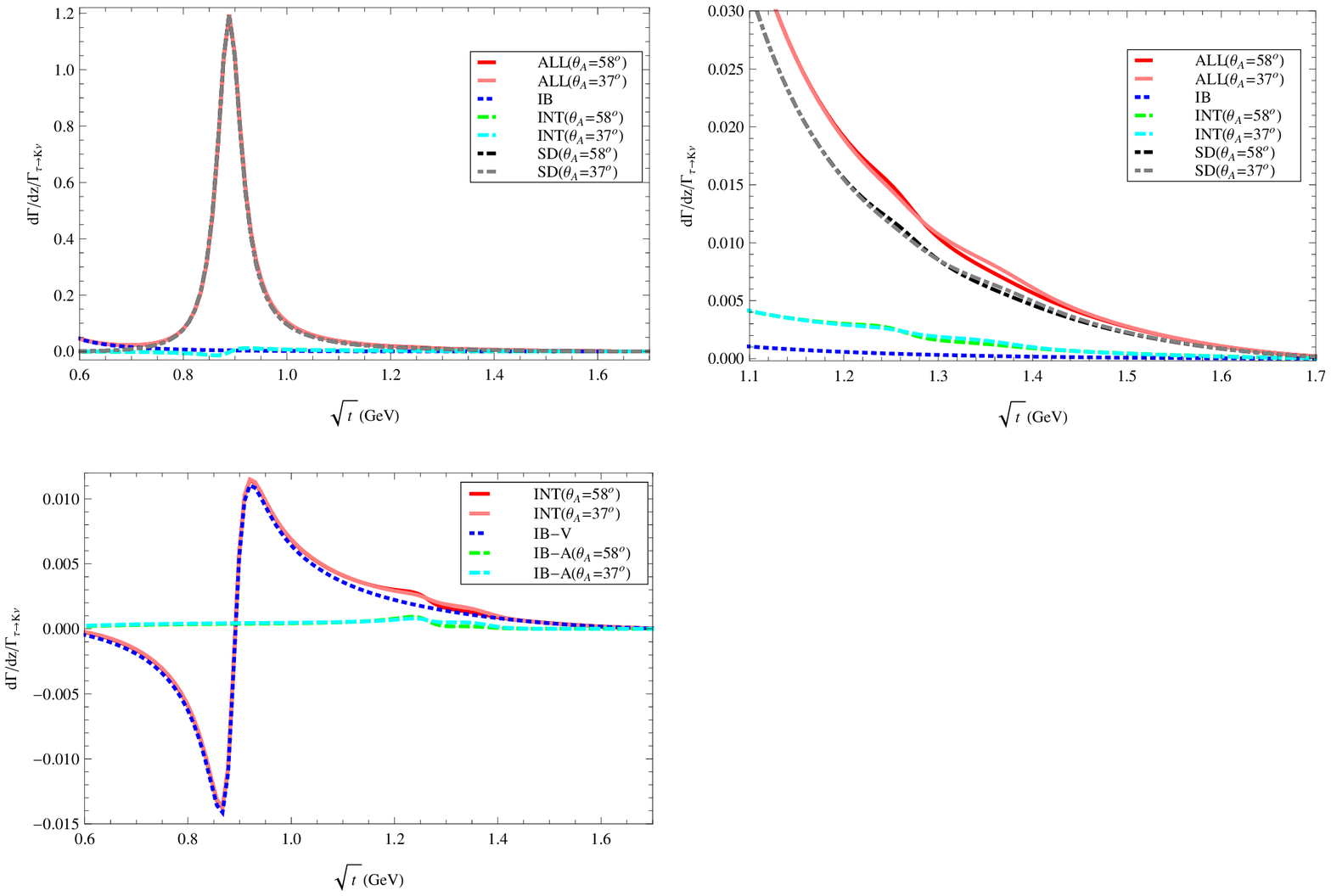}
\caption{ Differential decay width ( in units of $\Gamma_{\tau \to K \nu_\tau}$ ) of the process $\tau^-\rightarrow K^- \gamma \nu_\tau$
including all contributions as a function of $\sqrt t$ with $c_4=-0.07$. The top right plot is the close-up of the left one in the region of
$\sqrt t > 1.1$ GeV. The bottom left one is to show the compositions
of the interference parts. \label{fig.ktot4Zc4007}}
\end{center}
\end{figure}

Considering all the sources of uncertainty commented, we will content ourselves with giving our predictions for the case of
$c_4=0 \,,-0.07$ and $|\theta_A|=58^\circ \, , 37^\circ$. We first show the results of the decay rates in Table.\ref{Tabledifferentcontributionsk}.
The lesson we can learn from the numbers in Table \ref{Tabledifferentcontributionsk} is that the decay rate is sensitive to the value of $c_4$,
while different choices of $\theta_A$ barely influence the final results. In order to illustrate our predictions, we present the analogous plots
to those as we discussed in the $\tau^-\rightarrow  \pi^- \gamma \nu_\tau$ channel for the case $c_4=0$ and $|\theta_A|=37^\circ\,,58^\circ$
in Figs.\ref{fig.ktot4Xc40} and \ref{fig.ktot4Zc40}. For $c_4=-0.07$, we give the plots in Figs.\ref{fig.ktot4Xc4007} and \ref{fig.ktot4Zc4007},
where one can see that the $VV$ contribution from the $SD$ parts overwhelmingly dominates almost the whole spectrum.

\begin{table}[h!]
\begin{center}
\renewcommand{\arraystretch}{1.2}
\begin{tabular}{|c|c|c|c|c|}
\hline
&$x_0=0.0565\,,\, c_4=-0.07$ &$x_0=0.0565\,,\, c_4=0$  &$x_0=0.45\,,\, c_4=-0.07$  &$x_0=0.45\,,\, c_4=0$  \\
& $|\theta_A|=58^\circ (37^\circ)$& $|\theta_A|=58^\circ (37^\circ)$& $|\theta_A|=58^\circ (37^\circ)$ & $|\theta_A|=58^\circ (37^\circ)$ \\
\hline
$IB$&$3.64\cdot10^{-3}$&$3.64\cdot10^{-3}$&  $0.31\cdot10^{-3}$  &$0.31\cdot10^{-3}$\\
$IB-V$&$0.69\cdot10^{-3}$&$0.10\cdot10^{-3}$  &$0.83\cdot10^{-3}$  &$0.12\cdot10^{-3}$\\
$IB-A$&$0.22(0.25)\cdot10^{-3}$&$0.22(0.25)\cdot10^{-3}$ &$0.15(0.18)\cdot10^{-3}$   & $0.15(0.18)\cdot10^{-3}$\\
$VV$&$58.55\cdot10^{-3}$  &$1.30\cdot10^{-3}$  &$29.04\cdot10^{-3}$  &$0.66\cdot10^{-3}$\\
$VA$&$ \sim 0 (\sim 0) $&$ \sim 0 (\sim 0) $&$0.09(0.09)\cdot10^{-3}$&$0.01(0.01)\cdot10^{-3}$\\
$AA$&$0.13(0.16)\cdot10^{-3}$&$0.13(0.16)\cdot10^{-3}$ &$0.12(0.15)\cdot10^{-3}$  &$0.12(0.15)\cdot10^{-3}$\\
\hline
$ALL$&$63.23(63.29)\cdot10^{-3}$&$5.39(5.45)\cdot10^{-3}$&$30.54(30.60)\cdot10^{-3}$&$1.37(1.43)\cdot10^{-3}$\\
\hline
\end{tabular}
\caption{\small{Contributions of the different parts to the total rate in the decay $\tau^-\to K^-\gamma\nu_\tau$ (in units of $\Gamma_{\tau \to K \nu}$),
using two different cut-offs for the
photon energy: $E_\gamma=50$ MeV ($x_0=0.0565$) and $E_\gamma=400$ MeV ($x_0=0.45$) and also different values of the resonance couplings.
The numbers inside the parentheses denote the corresponding results with $|\theta_A|= 37^\circ$, while the other numbers are obtained with $|\theta_A|= 58^\circ$.   }}
\label{Tabledifferentcontributionsk}
\end{center}
\end{table}

\section{Conclusions} \label{Conclusions}

In this article we have studied the radiative one-meson decays of the $\t$: $\tau^-\rightarrow  (\pi/K)^- \gamma \nu_\tau$. We
have computed the relevant form factors for both channels and obtained the asymptotic conditions on the couplings imposed by the high-energy behaviour
of these form factors, dictated by $QCD$. The relations that we have found here are compatible with those found in other phenomenological applications
of the theory.

One of our motivations to examine these processes is that they have not been detected yet, according to naive estimates
or to Breit-Wigner parametrizations. We have checked the existing computations for the $IB$ part. Adding to it the $WZW$ contribution, that is the
$LO$ contribution in $\CPT$ coming from the $QCD$ anomaly, we have estimated the model independent contribution to both decays, that could be taken
as a lower bound. The values that we obtain for the $\pi$ channel are at least one order of magnitude above the already-observed $3K$ decay channel
even for a high-energy cut-off on the photon energy. In the $K$ channel, the model independent contribution gives a $BR$ larger than that of the $3K$
decay channel, as well. Only imposing a large cut-off on $E_\gamma$ one could understand that the latter mode has not been detected so far. We expect,
then, that upcoming measurements at $B$ and tau-charm factories will bring the discovery of these tau decay modes in the near future.

We do not have any free parameter in the $\tau^-\rightarrow  \pi^- \gamma \nu_\tau$ decay and that allows us to make a complete
study. Since the $IB$ contribution dominates, it will require some statistics to study the $SD$ effects. In this sense, the analysis of the $\pi$-photon spectrum
($t$-spectrum)
is more promising than that of the pure photon spectrum ($x$-spectrum), as we have shown. We are eager to see whether the discovery of this mode confirms our findings, since
we believe that the uncertainties of our study are small for this channel.

As expected, the higher mass of the Kaon makes easier the observation of $SD$ effects. However, there are several sources of uncertainty
in the $\tau^-\rightarrow  K^- \gamma \nu_\tau$ decay that prevent us from having a definitive prediction for this channel.
The most important one either
rises some doubts about the value of $c_4$, a parameter describing the $SU(3)$ breaking effect, obtained in Ref.~\cite{Dumm:2009kj} or on the sufficiency of
one multiplet of vector resonances to
describe this decay. We point out that the inclusion of the heavier multiplets of resonances will not only directly
give contributions to the spectra and the decay rates, but also influence the final results in an indirect way by
entering the resonance parameters given in Eq.(\ref{heconstraint}).
 As we have shown, the value of this coupling affects drastically the strength of the $VV$ (and thus the whole $SD$)
contribution. Besides, there is an uncertainty associated with the broad band of allowed values for $\theta_A$. However the $AA$ contribution
is anyway not important with respect to that on $c_4$. Even smaller is the error associated with the off-shell width behaviour
of the axial-vector neutral resonance with strangeness, $K_{1H,L}$. Since we have not calculated the relevant three-meson decay of the tau, we do
not have this expression within $R\chi T$ yet. We took a simple parametrization including the on-shell cuts corresponding to the decay chains
$K_{1H,L}\to (\rho K/K^* \pi)$. Since the effect of $c_4$ is so large, we expect that once this decay mode is discovered we will be able to bound this coupling.

As an application of this paper, we are working out \cite{GuoRoig} the consequences of our study in lepton universality tests through the ratios $\Gamma\left(
\tau^-\to\pi^-\nu_\tau \gamma\right)/\Gamma\left( \pi^-\to\mu^-\nu_\mu \gamma\right)$ and
$\Gamma\left( \tau^-\to K^-\nu_\tau \gamma\right)/\Gamma\left( K^-\to\mu^-\nu_\mu \gamma\right)$ that were considered in Refs.~
\cite{Decker:1993py,Sirlin:1981ie, Marciano:1988vm} in different frameworks. The ratio between the decays in the denominators within $\CPT$ was studied
in \cite{Cirigliano:2007xi, Cirigliano:2007ga} and the radiative pion decay has been investigated within $\RCT$ in \cite{Mateu:2007tr} recently.\\

\section*{Acknowledgments}
We are indebted to Jorge Portol\'es for reading our draft for this article and making useful comments and criticisms. We wish to thank
Olga Shekhovtsova and Zbigniew Was for their interest in our work and useful discussions on this topic. Z.~H.~Guo and P.~Roig ackowledge
the financial support of the Marie Curie ESR Contract (FLAVIAnet). This work has been supported in part by the EU MRTN-CT-2006-035482 (FLAVIAnet) and
by the Spanish Consolider-Ingenio 2010 Programme CPAN (CSD2007-00042).

\appendix

\section{Off-shell widths of the intermediate resonances}\label{appendix-width}
The off-shell widths of the resonances used in our numerical discussion are taken from Refs. \cite{Guo:2008sh, Dumm:2009va, GomezDumm:2000fz}.
The off-shell width of the $\rho(770)$ is
\bqa
\Gamma_\rho(s) = \frac{s\, M_\rho}{96 \pi F^2} \bigg[ \sigma_{\pi\pi}^3(s)  + \frac{1}{2}\sigma_{KK}^3(s) \bigg]\,,
\eqa
where
\begin{equation}
 \sigma_{P Q}(s)\,=\,\frac{1}{s}\,\sqrt{\left[s-(m_P+m_Q)^2\right]\left[s-(m_P-m_Q)^2\right]}\,\,\theta\left[ s-(m_P+m_Q)^2\right]\,.
\end{equation}
$\theta(x)$ is the standard unit step function.

The off-shell width of the $K^*(892)$ is
\bqa
\Gamma_{K^*}(s) = \frac{s\, M_{K^*}}{128 \pi F^2} \bigg[ \sigma_{\pi K}^3(s)  + \sigma_{\eta K}^3(s) \bigg]\,,
\eqa
while that of the $K_{1A}$ is
\begin{equation}
 \Gamma_{K_{1A}}(s)\,=\,\Gamma_{K_{1A}}(M_{K_{1A}}^2)\,\frac{s}{M_{K_{1A}}^2} \bigg[ \frac{\sigma^3_{ K^* \pi}(s)+\sigma^3_{ \rho K}(s)}{\sigma^3_{K^* \pi}
 (M_{K_{1A}}^2)+\sigma^3_{ \rho  K}(M_{K_{1A}}^2)} \bigg]\,.
\end{equation}

The $a_1(1260)$ off-shell width is
\begin{equation} \label{eq:Gamma_a1_tot}
\Gamma_{{\rm a}_1}(Q^2) \ = \ \Gamma_{{\rm a}_1}^{3\pi}(Q^2)\,  \theta(Q^2-9m_{\pi}^2) \
+ \ \Gamma_{{\rm a}_1}^{KK\pi}(Q^2) \, \theta(Q^2-(2 m_K+m_{\pi})^2)  \ ,
\end{equation}
where
\begin{equation} \label{eq:Gamma_a1_pimod}
\Gamma_{{\rm a}_1}^{3\pi,KK\pi}(Q^2) \ = \ \frac{-S}{192(2\pi)^3 F_A^2 M_{{\rm a}_1}}\;
\left( \frac{M_{{\rm a}_1}^2}{Q^2}-1 \right)^2 \; \int \mathrm{d}s\, \mathrm{d}t\;
T^{3\pi,KK\pi\mu}_{1^+}\; T^{3\pi,KK\pi\ast}_{1^+\mu} \ .
\end{equation}
Here $\Gamma_{{\rm a}_1}^{3\pi}(Q^2)$ recalls the three pion contributions and
$\Gamma_{{\rm a}_1}^{KK\pi}(Q^2)$ collects the contributions of the $KK\pi$ channels.
In Eq.~(\ref{eq:Gamma_a1_pimod}) the symmetry factor $S = 1/n!$ recalls
the case with $n$ identical particles in the final state. The explicit expressions for
$T^{3\pi,KK\pi\mu}_{1^+}$ can be found in Ref.\cite{Dumm:2009va}.
We stress that the on-shell width, $\Gamma_{{\rm a}_1}(M_{{\rm a}_1}^2)$, is
 a prediction and not a free parameter.

\section{Numerical inputs}\label{appendix-input}
In the numerical discussion, unless a specific statement is given, we use the values given in Ref.\cite{PDG08}.
For the pion, kaon and $K^*(892)$, we use the masses of the charged particles throughout, i.e.,
\bqa
m_\pi = 139.6\, {\rm MeV} \,,\quad  m_K = 493.7 \, {\rm MeV}\,,\quad  M_{K*} = 891.7\, {\rm MeV}\,.
\eqa
For the mass of $a_1(1260)$, we use the result of $M_{a_1} = 1120$ MeV from Ref.\cite{Dumm:2009va}, which has taken the off-shell width effect into account.

For the resonance couplings, once we have the value of the pion decay constant $F$ in the chiral limit, we can determine
all of the others except $c_4$ through Eq.(\ref{heconstraint}). For the value of $F$, we use $F=90$ MeV. The physical pion and kaon decay constants
we are using are
\bqa
F_\pi = 92.4 \,{\rm MeV}\,, \quad F_K= 113.0 \,{\rm MeV}\,.
\eqa

\end{document}